\documentclass[twocolumn]{aastex62}
\pdfoutput=1 %for arXiv submission
\usepackage[version=4]{mhchem}
\usepackage{amsmath}
\usepackage{multirow}
\usepackage{hyperref}
\usepackage[utf8]{inputenc}
\usepackage{comment}
\usepackage{amsmath}

\newcommand{\revision}[1]{\textcolor{black}{#1}}
\newcommand{\revisiontwo}[1]{\textcolor{black}{#1}}

\graphicspath{{./}{figures/}}

\received{TBD}
\revised{TBD}
\accepted{TBD}
\shorttitle{Chemical disequilibrium fingerprints on planetary spectra}
\shortauthors{Molaverdikhani et al.}

\begin{document}

\title{From cold to hot irradiated gaseous exoplanets: Fingerprints of chemical disequilibrium in atmospheric spectra}

\correspondingauthor{Karan Molaverdikhani}
\email{Karan@mpia.de}

\affiliation{Max Planck Institute for Astronomy, Königstuhl 17, 69117 Heidelberg, Germany}

\author{Karan Molaverdikhani}
\affiliation{Max Planck Institute for Astronomy, Königstuhl 17, 69117 Heidelberg, Germany}

\author{Thomas Henning}
\affiliation{Max Planck Institute for Astronomy, Königstuhl 17, 69117 Heidelberg, Germany}

\author{Paul Molli\`ere}
\affiliation{Sterrewacht Leiden, Huygens Laboratory, Niels Bohrweg 2, 2333 CA Leiden, The Netherlands}

%%%%%%%%%%%%%%%%%%%%%%%%%%%%%%%%%%%%%

\begin{abstract}

Almost all planetary atmospheres are affected by disequilibrium chemical processes. In this paper we \revision{introduce  our recently developed Chemical Kinetic Model (\texttt{ChemKM}). We show that the results of our HD\;189733b model are in good agreement} with previously published results, except at $\mu$bar regime, where molecular diffusion and photochemistry are the dominant processes. We thus recommend careful consideration of these processes when abundances at the top of the atmosphere are desired. We also propose a new metric for a quantitative measure of quenching levels. By applying this metric, we find that quenching pressure decreases with the effective temperature of planets, but it also varies significantly with other atmospheric parameters such as [Fe/H], log(g), and C/O. In addition, we find that the ``Methane Valley'', a region between 800 and 1500\;K where \revision{above a certain C/O threshold value} a greater chance of \ce{CH4} detection is expected, still exists after including the vertical mixing. The first robust \ce{CH4} detection on an irradiated planet (HD\;102195b) places this object within this region; supporting our prediction. \revision{We also investigate the detectability of disequilibrium spectral fingerprints by JWST, and suggest focusing on the targets with T\textsubscript{eff} between 1000 and 1800\;K, orbiting around M-dwarfs, having low surface gravity but high metallicity and a C/O ratio value around unity.} Finally, constructing Spitzer color-maps suggests that the main two color-populations are largely insensitive to the vertical mixing. Therefore any deviation of observational points from these populations are likely due to the presence of clouds and not disequilibrium processes. However, some cold planets (T\textsubscript{eff}$<$900\;K) with very low C/O ratios ($<$0.25) show significant deviations; making these planets interesting cases for further investigation.

\end{abstract}


%%%%%%%%%%%%%%%%%%%%%%%%%%%%%%%%%%%%%%%%%%%%%%%%%%%%%%%%%%%
%%%%%%%%%%%%%%%%%%%%%%%%%%%%%%%%%%%%%%%%%%%%%%%%%%%%%%%%%%%
%%%%%%%%%%%%%%%%%%%%%%%%%%%%%%%%%%%%%%%%%%%%%%%%%%%%%%%%%%%
%\tableofcontents

%%%%%%%%%%%%%%%%%%%%%%%%%%%%%%%%%%%%%%%%%%%%%%%%%%%%%%%%%%%
\section{Introduction} \label{sec:intro}

The atmospheric composition of solar system planets is not at their thermochemical equilibrium state, mostly due to the irradiation by the Sun and mixing through atmospheric transport and turbulence. These effects are so pronounced that even the earliest atmospheric models of these objects include disequilibrium chemistry (e.g. review articles on the early modern models of Venus \citep{noll_models_1972}, Earth \citep{lou_models_1973}, Mars \citep{noll_models_1974}, Jupiter \citep{danielson_models_1968}, Saturn \citep{divine_models_1972}, Titan \citep{divine_titan_1974}, Uranus, and Neptune \citep{encrenaz_observational_1974}).

With the discovery of Hot Jupiters \citep{mayor_jupiter-mass_1995}, however, the assumption of thermochemical equilibrium resurfaced as a first-order estimation of their atmospheric properties thanks to their high temperatures \citep[e.g.][]{burrows_chemical_1999,lodders_atmospheric_2002,fortney_comparative_2005}. But further studies suggested that the non-thermal processes could potentially alter their chemical composition at the photospheric levels; predominantly because of the intense UV irradiation and strong atmospheric mixing \citep[e.g.][]{seager_exoplanet_2005,madhusudhan_temperature_2009,stevenson_possible_2010,moses_disequilibrium_2011,agundez_impact_2012,venot_chemical_2012,hu_photochemistry_2012,hu_photochemistry_2013,hebrard_photochemistry_2013,hu_photochemistry_2014,miguel_effect_2014,zahnle_methane_2014,drummond_effects_2016,tsai_vulcan:_2017,wang_modeling_2017,blumenthal_comparison_2018,zhang_global-mean_2018,changeat_towards_2019} or other processes such as lightning \citep{helling_lightning_2019} and cosmic rays \citep{rimmer_ionization_2013}.

\citet{madhusudhan_high_2011} performed an extensive retrieval analysis of the GJ\;436b spectrum observed by \citet{stevenson_possible_2010} and concluded that the methane abundance paucity cannot be explained by thermoequilibrium chemistry. While the lack of a robust methane detection in the spectra of irradiated gaseous planets \citep[e.g.][]{brogi_rotation_2016,brogi_exoplanet_2018,pino_combining_2018,alonso-floriano_multiple_2019} could be an indication of disequilibrium chemistry, there is no \revision{large-scale} systematic investigation of how disequilibrium processes change the abundance of methane (or any other major opacity compound) at photospheric pressures \revision{($\sim$1\;bar to $\sim$1\;$\mu$bar)}, and how this manifests itself in the atmospheric spectra. \revision{\citet{zahnle_methane_2014} and \citet{miguel_exploring_2014} both perform such studies however over a limited parameter space.} The complexity of photochemical models makes the \revision{large-scale} simulations computationally expensive, and thus such investigations require code optimization. In addition, the models must be flexible and generic enough to perform efficiently over a range of atmospheric conditions and compositions.

We introduce a new chemical kinetic model (\texttt{ChemKM}), which is both fast and generic. We employ \texttt{ChemKM} to perform an extensive parameter study over a broad range of atmospheric conditions. In total, we calculate 84,672 full-network chemical kinetic models with more than 100 reactants and 1000 reactions, to study how the variation of major opacity sources (e.g. \ce{H2O}, \ce{CH4}, \ce{CO}, and \ce{CO2}) due to disequilibrium chemistry affect the spectra. We also present a case study of HD\;189733b, a well-studied exoplanet, for benchmarking.

\revision{In what follows, we describe our \revision{chemical kinetic} model and the parameter space that we have explored. In  \hyperref[sec:HD189]{Section}\;\ref{sec:HD189}, we present the results of our case study for HD\;189733b and benchmark it against the model developed by \citet{venot_photochimie_2012}. In the rest of \hyperref[sec:results]{Section}\;\ref{sec:results}, we present the results of our parametric study and check the validity of our proposed classification scheme for irradiated planetary spectra (see \citet{molaverdikhani_cold_2019} for more details on the classification under cloud-free equilibrium chemistry conditions). 
In \hyperref[subsec:Quenching]{Section}\;\ref{subsec:Quenching}, we report the dependency of quenching point on atmospheric parameters, and in \hyperref[subsec:JWST]{Section}\;\ref{subsec:JWST} we investigate the detectability of disequilibrium processes by JWST. Finally, in \hyperref[subsec:Classification]{Section}\;\ref{subsec:Classification} we discuss Spitzer transmission and emission color diagrams variation from thermochemical equilibrium models due to the vertical mixing, and its observability. We summarize and conclude our results and findings in \hyperref[sec:discussion]{Section}\;\ref{sec:discussion}.}

%>>>>>>>>>>>>>>>>>>>
%\note{Show the initial condition and model setup of each figure/model in it with symbols or other legible methods, e.g. IC: elemental abundance; Setup: Photochemistry (radiation/star icon on  the figures) e.g. IC: Photo-diff (radiation and wind icons) eq.; Setup:influx/condensation (waterfall/ice icons)}

%%%%%%%%%%%%%%%%%%%%%%%%%%%%%%%%%%%%%%%%%%%%%%%%%%%%%%%%%%%
\section{Methods} \label{sec:method}
%In order to study the effects of disequilibrium chemistry on the atmospheric spectra of irradiated planets, the model must be flexible enough to cover a variety of physical ranges, e.g. from cold to hot environments, and yet be computationally fast.

%%%%%%%%%%%%%%%%%%%%%%%%%%%%%%%%%%%%%%%%%%%%%%%%%%%%%%%%%%%
\subsection{\texttt{ChemKM}: The Chemical Kinetic Model} \label{subsec:ChemKM}

To study the effects of disequilibrium chemistry on the composition and atmospheric spectra of irradiated exoplanets we developed a 1D Chemical Kinetic Model (\texttt{ChemKM}). The abundances of atmospheric constituents are governed by numerically solving the equations of chemical kinetics (also known as the altitude-dependent continuity–diffusion equation) which describe the formation and destruction of species, \hyperref[eq:continuity]{equation} \ref{eq:continuity} \citep[e.g.][]{allen_vertical_1981}:

\begin{equation} 
\frac{\partial n_i}{\partial t} = P_i - L_i -\frac{\partial \Phi_i}{\partial z},
\label{eq:continuity}
\end{equation}

where $n_i$ is the number density (cm$^{-3}$), $P_i$ is the chemical production rate (cm$^{-3}$ s$^{-1}$), $L_i$ is the chemical loss rate (cm$^{-3}$ s$^{-1}$), and $\Phi_i$ is the net vertical flux (cm$^{-2}$ s$^{-1}$) of species $i$ at altitude z. These quantities are all functions of time $t$ and altitude $z$ (or alternatively pressure). The vertical flux transport terms include molecular and eddy diffusion, $\Phi_i$=$\Phi_{i,mol}$+$\Phi_{i,eddy}$ where the latter is commonly parameterized by an eddy diffusion coefficient K\textsubscript{zz}. The vertical flux transport terms can be estimated through  \hyperref[eq:moldiff]{equations} \ref{eq:moldiff} and \ref{eq:eddydiff}, respectively (see e.g. \citet{moses_i._1991} and references therein).

\begin{equation} 
\Phi_{i,mol}=-n_T D_i \Big( \frac{\partial f_i}{\partial z} - \frac{f_i}{H_a} + \frac{f_i}{H_i} + \frac{\alpha_if_i}{T}\frac{dT}{dz} \Big)
\label{eq:moldiff}
\end{equation}

\begin{equation} 
\Phi_{i,eddy}=-n_T K_{zz} \Big( \frac{\partial f_i}{\partial z} \Big),
\label{eq:eddydiff}
\end{equation}

where K\textsubscript{zz} is the eddy diffusion coefficient (cm$^2$ s$^{-1}$), $D_i$ is the molecular diffusion coefficient (cm$^2$ s$^{-1}$), $n_T$ is the total number density, $f_i$ is the mixing ratio of species $i$ and is defined as $f_i=n_i/n_T$, $H_a$ is the mean scale height of the atmosphere, $H_i$ is the scale height of species i, $T$ is the temperature (K), and $\alpha_i$ is the thermal diffusion factor of species $i$.

We use Lennard-Jones calculations for the estimation of molecular diffusivity of species. An updated list of  Lennard-Jones potentials can be obtained from Appendix-B of \citet{poling_properties_2000}. Note that $D_i$ scales inversely with pressure and will become important at very low pressures. Therefore this term's contribution is no longer negligible at the pressures probed by transmission observations, \revision{i.e. above $\sim$1\;bar}. The effect becomes more pronounced when the atmosphere is probed by high resolution spectroscopy. We discuss this in more detail in \hyperref[sec:HD189]{Section}\;\ref{sec:HD189}.

One approach to estimate K\textsubscript{zz} is using general circulation models (GCMs) to calculate wind velocity fields (e.g. \citet{moses_disequilibrium_2011}) or to compute the advection of passive tracers \citep{parmentier_3d_2013}. Due to large uncertainties in these approaches an alternative method could be to treat the eddy diffusion coefficient as a free parameter in the model \citep[e.g.][]{miguel_exploring_2014}. We use the latter approach, in our parametric study.

\revision{For an overview of our methodology, numerical solution, and chemical networks see \hyperref[subsec:methodology]{Appendix}\;\ref{subsec:methodology}. In \hyperref[sec:verification]{Appendix}\;\ref{sec:verification},} we present the results of verification of our model under different conditions such as thermochemical equilibrium or by including molecular and eddy diffusion, photochemistry, condensation, and setting up atmospheric influxes and different boundary conditions.

%%%%%%%%%%%%%%%%%%%%%%%%%%%%%%%%%%%%%%%%%%%%%%%%%%%%%%%%%%%
\subsection{\revision{The temperature structures for the parametric study}} \label{subsec:petitCODE}

In \hyperref[sec:results]{Section}\;\ref{sec:results}, we study a broad range of parameter space, based on a large grid of cloud-free atmospheric models. We use the self-consistently calculated temperature-abundances profiles as the input of our chemical kinetic model (\texttt{ChemKM}) to calculate the effect of vertical mixing on the chemical composition of the atmospheres. \revision{The input profiles were calculated by the petitCODE \citep{molliere_model_2015,molliere_observing_2017}. We then use petitRADTRANS to calculate the spectra of atmospheres at chemical disequilibrium} to investigate how the quenching point changes with atmospheric parameters, and whether the classification scheme proposed in \citep{molaverdikhani_cold_2019} would still hold after introducing vertical mixing. Here we briefly review the properties of the grid and the range of investigated parameters.

In the first paper of this series \citep{molaverdikhani_cold_2019} we calculated a large grid of 28,224 self-consistent cloud-free atmospheric models using the petitCODE. The petitCODE is able to calculate planetary atmospheric temperature profiles, chemical abundances, and emergent and transmission spectra and it assumes radiative-convective and thermochemical equilibrium in a 1D setup. The stellar effective temperature, stellar radius, planetary effective temperature or distance, planetary internal temperature, planetary radius, planetary mass or alternatively its surface gravity and irradiation treatment must be provided.

We considered five factors as the free parameters in our petitCODE grid models: planetary effective temperature (T\textsubscript{eff}), surface gravity (log(g)), metallicity ([Fe/H]), carbon-to-oxygen ratio (C/O) and spectral type of the host star. The lower limit of T\textsubscript{eff} was set to 400\;K to avoid non-negligible contributions of the interior temperature (all models have T\textsubscript{int} = 200\;K) into the atmospheric properties, and the upper limit was chosen to be 2600\;K to avoid highly irradiated regions where the heat redistribution becomes inefficient and the planetary average irradiation treatment might become invalid. The surface gravity spans from 2.0 to 5.0 to broadly cover possible surface gravity values. Metallicity ranges from -1.0 to 2.0 with an increment of 0.5. The stellar spectral types were chosen to be M5, K5, G5 and F5. For C/O we selected irregular parameter steps ranging from 0.25 to 1.25 with smaller steps around unity to capture possible transitions from water- to methane- dominated atmospheres as predicted by \citet{madhusudhan_c/o_2012}, reported by \citet{molliere_model_2015}, and discussed in detail in \citet{molaverdikhani_cold_2019}. We then add K\textsubscript{zz} as a new dimension to be explored; more discussion and results in \hyperref[sec:results]{Section}\;\ref{sec:results}. Planetary spectra are calculated with petitRADTRANS \citep[][]{molliere_petitradtrans:_2019}.

%%%%%%%%%%%%%%%%%%%%%%%%%%%%%%%%%%%%%%%%%%%%%%%%%%%%%%%%%%%

\section{Results} \label{sec:results}
\subsection{Case study: HD\;189733b} \label{sec:HD189}

%%%%%%%%%%%%%%%%%%%%%
\begin{figure*}
\includegraphics[width=\textwidth]{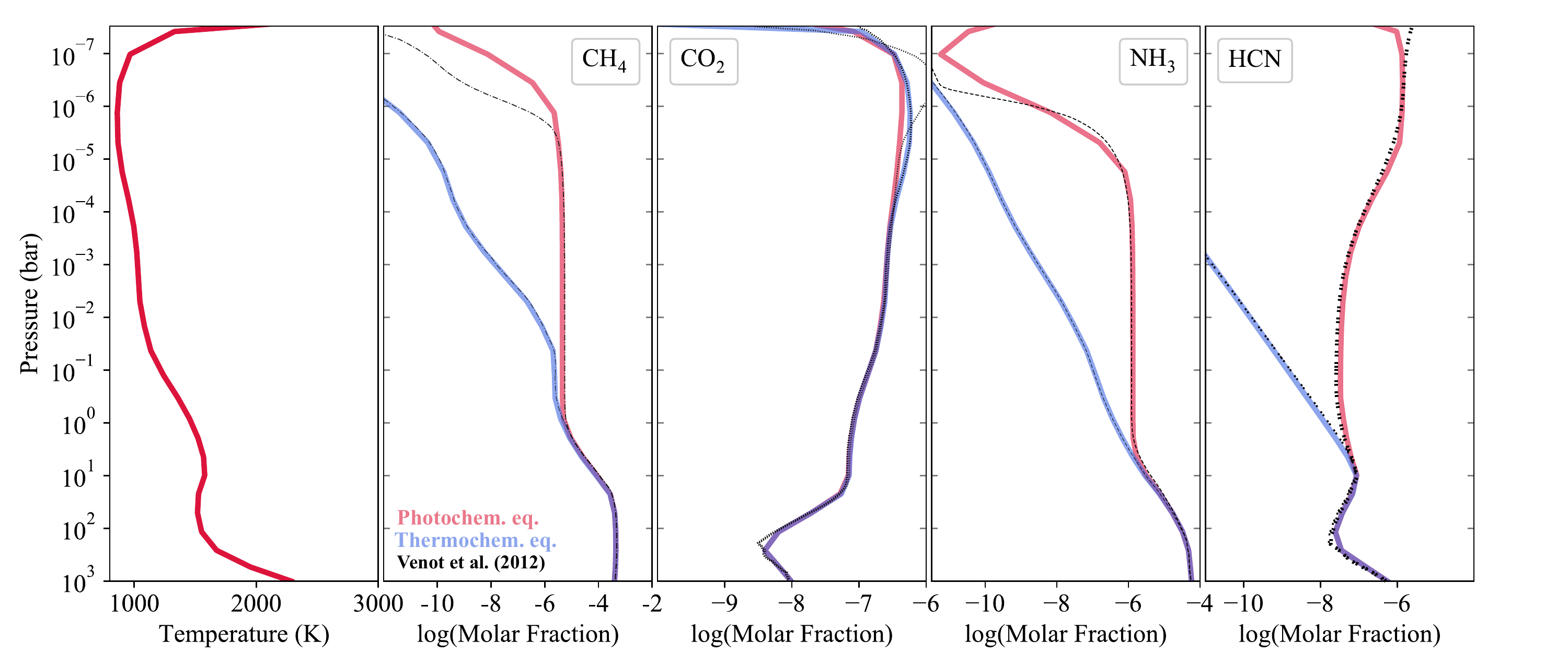}
\caption{HD\;189733b photochemical model comparison. {\bf Left)} The temperature structure. {\bf Rest of panels)}  \texttt{ChemKM}'s calculated abundances of \ce{CH4}, \ce{CO2}, \ce{NH3}, \ce{HCN}, and \ce{H} at thermochemical equilibrium (solid blue lines) and photo-diffusion equilibrium (solid red lines), compared with the results of \citet{venot_chemical_2012} (black lines). The results are in good agreement except at and above the $\mu$bar regime, where photolysis reactions and molecular diffusion are the dominant processes. Hence a careful implementation of these processes is necessary if TOA abundances are desired. \label{fig:HD198}}
\end{figure*}
%%%%%%%%%%%%%%%%%%%%%

HD\;189733b is one of the most studied exoplanets so far \citep[e.g.][]{moses_disequilibrium_2011} and hence represent a proper case for benchmarking. \citet{moses_disequilibrium_2011}
provided the thermal structure and the K\textsubscript{zz} profile for this planet, which has been used by several studies to perform benchmark calculations \citep[e.g.][]{venot_chemical_2012,moses_chemical_2014,drummond_effects_2016,tsai_vulcan:_2017}. Thus we use the same set of inputs to compare the outcome of our HD\;189733b model with previously published results.

To set up the model, we use \citet{venot_chemical_2012}'s full kinetic network and an updated version of \citet{hebrard_neutral_2012}'s UV absorption cross-sections and branching yields. For UV irradiation at TOA, both \citet{moses_disequilibrium_2011} and \citet{venot_chemical_2012} used $\epsilon$ Eridani as a proxy of HD\;189733 due to their similarities in spectral type, age, and metallicity. Lack of high quality data at shorter wavelengths led them to use only a portion of $\epsilon$ Eridani's spectrum (\citet{moses_disequilibrium_2011} used $\epsilon$ Eridani's data in the range of 115\;nm to 283\;nm and \citet{venot_chemical_2012} uses data between 90\;nm and 330\;nm). They combined other datasets and models to extend the spectrum. We use the most recent measurements of $\epsilon$ Eridani as a full and coherent spectrum from X-ray to optical \revisiontwo{to take a more consistent approach}. The data were obtained from the MUSCLES database \citep{france_muscles_2016}. We initialize the atmospheric composition at its thermochemical equilibrium state and let the model reach its steady state after an integration time of 10$^9$\;sec. 

\citet{venot_chemical_2012} and \citet{moses_chemical_2014} reported the importance of kinetic network and its effect on the quenching level and quenching abundances. Since we opted for the usage of \citet{venot_chemical_2012}'s network, we compare our results with their findings. Our thermochemical equilibrium abundances (blue lines in \hyperref[fig:HD198]{Figure}\;\ref{fig:HD198}) are almost identical to the \citet{venot_chemical_2012}'s results. Similarly, our disequilibrium abundances (red lines in \hyperref[fig:HD198]{Figure}\;\ref{fig:HD198}) are almost identical, except in the $\mu$bar regime. \revisiontwo{By using the same stellar flux as \citet{venot_chemical_2012} and finding no significant abundance variation, we rule out the role of stellar flux as a cause of discrepancies at high altitudes.} Therefore, these subtle differences between HD\;189733b photochemical models are likely caused by different molecular diffusion implementation and different photolysis reactions. As briefly mentioned in \hyperref[subsec:ChemKM]{Section}\;\ref{subsec:ChemKM}, this might have significant consequences in the interpretation of high-resolution spectra.

%Free high-resolution retrievals (with a minimum set of physical assumptions in their forward model) can potentially retrieve abundances, temperature and pressures at TOA. But to connect these to deeper pressures we have to have a better understanding of the physics and chemistry at all pressures. Hence it is necessary to make the models as comprehensive as possible. Obtaining low/mid resolution spectra of the same planet can also provide complementary information on the continuum, which represents the spectral signatures arising from the deeper parts of the atmosphere. Hence, a joint analysis of low-high resolution, accompanied with a comprehensive physico-chemical model, could act as a powerful tool to tackle the issue.

%In addition to the kinetic networks, different thermochemical coefficients, photolysis reactions, discretization of diffusion term, and even the vertical mesh could result in different outcomes. 

%%%%%%%%%%%%%%%%%%%%%%%%%%%%%%%%%%%%%%%%%%%%%%%%%%%%%%%%%%%
\subsection{\revision{Methane depletion in GJ\;436b}} \label{sec:GJ436b}

We discussed that while the atmospheric composition of deeper regions of gaseous planets tend to remain at thermochemical equilibrium, their photospheres are usually prone to disequilibrium processes. This effect may have its fingerprint on the planets' atmospheric spectra, but its detection is not a trivial task with the current observational facilities.

In a well-known case, \citet{stevenson_possible_2010} reported \ce{CH4} depletion inferred from the thermal emission of the GJ\;436b dayside (a Class-I planet with T$\sim$700\;K). \revision{Several studies} suggested that this \ce{CH4} deficiency, relative to its thermochemical equilibrium predictions, could be explained by disequilibrium processes such as diffusion and polymerization of \ce{CH4} \citep[see e.g.][]{stevenson_possible_2010, madhusudhan_high_2011,line_thermochemical_2011}. \citet{madhusudhan_high_2011} performed a detailed retrieval analysis of the spectrum and concluded that the atmosphere maintains a possible high metallicity (10$\times$ solar) and a vertical mixing with K\textsubscript{zz}$\approx$10$^6$-10$^7$\;(cm$^2$ s$^{-1}$). They constrained these values by assuming a suite of parametric TP structures. Under thermochemical equilibrium conditions, such temperature profiles could result in a monotonically decreasing \ce{CH4} abundance at the upper portion of the atmosphere. If the mixing timescale is much shorter than the kinetic timescales at those pressures, then the vertical mixing could ``transport'' these deeper/lower abundances and bring them to the upper levels. Hence a \ce{CH4} deficiency at TOA could occur under these conditions. Such abundance variation could change the measured flux by $Spitzer$ IRAC instrument in particular the 3.6\;$\mu$m channel \citep[][]{stevenson_possible_2010}.

\begin{figure*}
\includegraphics[width=\textwidth]{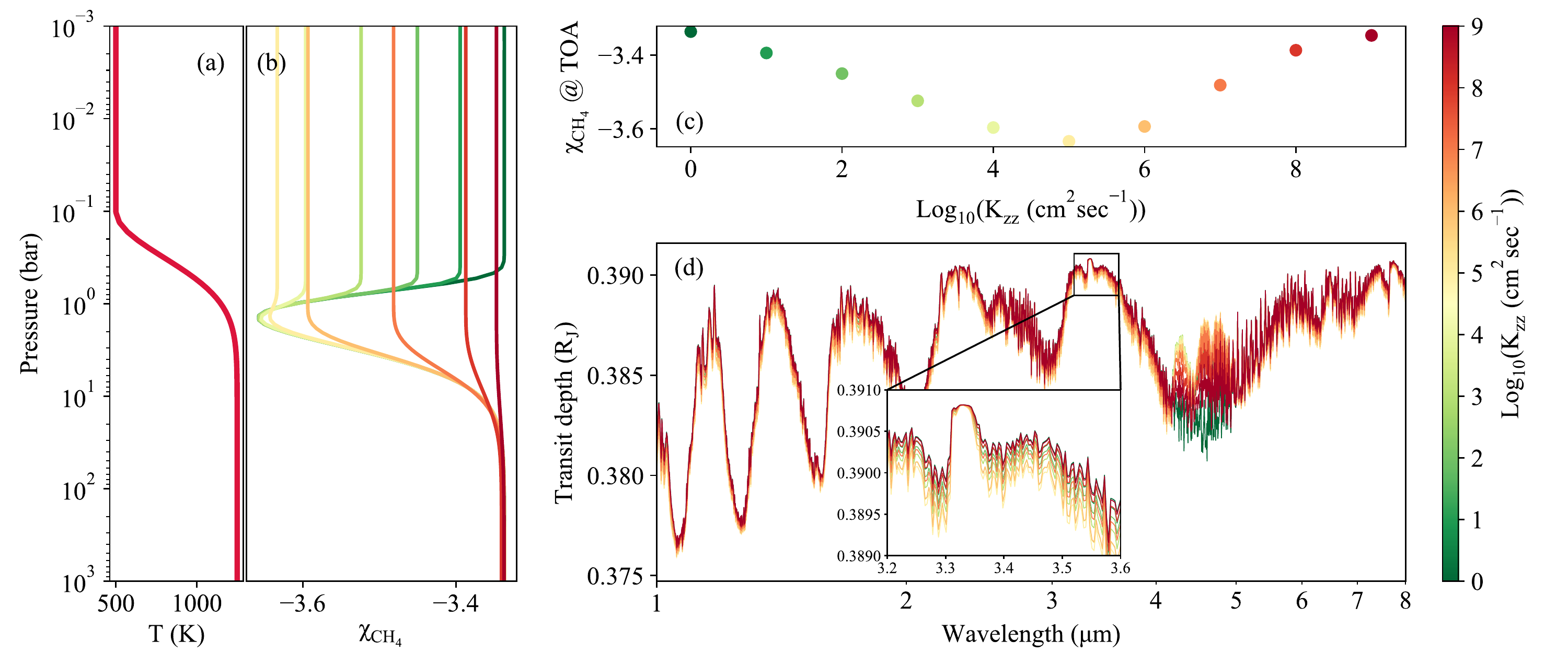}
\caption{Effect of vertical mixing on \ce{CH4} abundances at the top of the atmosphere. {\bf (a)} A simple parametric GJ\;436b$-$like TP structure, adapted from \citet{madhusudhan_high_2011}. {\bf (b)} \ce{CH4} abundance profiles caused by different K\textsubscript{zz} values. Stronger mixing causes a deeper quenching level, but does not guarantee a monotonic abundance variation. {\bf (c)} \ce{CH4} abundance at TOA as a function of K\textsubscript{zz} in this particular case. {\bf (d)} Variation of the transmission spectrum as a function of K\textsubscript{zz}. In this example, variation of \ce{CH4} abundance at TOA appears as a slight shift in the transmission spectrum. The \ce{CO2} and CO spectral features, between 4$-$5\;$\mu$m, varies as well; resulting in higher \ce{CO2} and CO molar fractions at TOA when the mixing is in action.} \label{fig:CH4depletion}
\end{figure*}

To illustrate how such an abundance deficiency occurs, we setup a simple model by selecting a parametric TP structure from \citet{madhusudhan_high_2011},  \hyperref[fig:CH4depletion]{Figure}\;\ref{fig:CH4depletion} (a). Here, we do not intend to perform a retrieval to reproduce the \citet{madhusudhan_high_2011}'s results, but rather it is a demonstration of abundance variation at TOA due to imposing the vertical mixing. We assume 500\;K and 1500\;K for the temperature of upper and deeper regions respectively, with solar metallicity and C/O ratio for the bulk composition of the atmosphere. With this setup, methane's thermochemical equilibrium abundance at TOA is around 4.6$\times$10$^{-4}$. By increasing the vertical mixing, the atmosphere quenches at deeper levels, see \hyperref[fig:CH4depletion]{Figure}\;\ref{fig:CH4depletion} (b). The \ce{CH4} abundance at TOA decreases as K\textsubscript{zz} increases, as long as quenching occurs at pressures smaller than $\sim$1\;bar. Any K\textsubscript{zz} higher than 10$^5$\;cm$^2$ s$^{-1}$, however, mixes the deeper levels where \ce{CH4} has higher abundances; causing to an enhancement of \ce{CH4} abundance at TOA. \hyperref[fig:CH4depletion]{Figure}\;\ref{fig:CH4depletion} (c) illustrates the variation of \ce{CH4} at TOA, due to the change of vertical mixing, and a minimum \ce{CH4} abundance in the case of K\textsubscript{zz}$=$10$^5$\;cm$^2$ s$^{-1}$. Here we only showed the results of \ce{CH4} abundance, however, almost all reagents respond to the variation of vertical diffusion; although differently. Their collective variation could affect the transmission spectrum of the planet as can be seen in \hyperref[fig:CH4depletion]{Figure}\;\ref{fig:CH4depletion} (d).

%%%%%%%%%%%%%%%%%%%%%
%%%%%%%%%%%%%%%%%%%%%

This non-linear atmospheric response is likely to be more pronounced for hotter planets, e.g. Class-II, where the abundance of methane and water (as two of the major opacity sources in the planetary atmospheres) are more sensitive to the variation of C/O ratio at their photospheres due to partial evaporation of condensates.

\revision{We expand these results by conducting} an extensive survey \revision{on the effect of} vertical mixing in the spectra of our self-consistently calculated cloud-free grid of models (see \hyperref[subsec:petitCODE]{Section}\;\ref{subsec:petitCODE} and \citep[][]{molaverdikhani_cold_2019}). By using these TP structures as \texttt{ChemKM}'s input and assuming three values for K\textsubscript{zz} (10$^6$, 10$^9$, and 10$^{12}$\;cm$^2$ s$^{-1}$) we calculate the atmospheric abundances of these planets at their diffusion equilibrium. \revision{The grid of models and their emission and transmission spectra will be publicly available.\footnote{www.mpia.de/homes/karan}} As the first step, we determine the dependency of quenching levels to the atmospheric parameters.

%%%%%%%%%%%%%%%%%%%%%%%%%%%%%%%%%%%%%%%%%%%%%%%%%%%%%%%%%%%
\subsection{Parametric study: Quenching levels}  \label{subsec:Quenching}

The atmospheric diffusion not always results in a constant ``quenched abundance'' above the ``quenching level'' (see, e.g., \hyperref[fig:reversreac1Ddiff]{Figure}\;\ref{fig:reversreac1Ddiff}, and \hyperref[subsec:eddy]{Section}\;\ref{subsec:eddy} for an overview). Therefore, this definition of quenching level (and consequently constant quenched abundance) is not general. Here we propose a parameter, the Coefficient of variation (CV), for a quantitative definition of quenching levels based on the deviation of abundances from their thermochemical equilibrium values at each pressure level.

\begin{figure}
\includegraphics[width=\columnwidth]{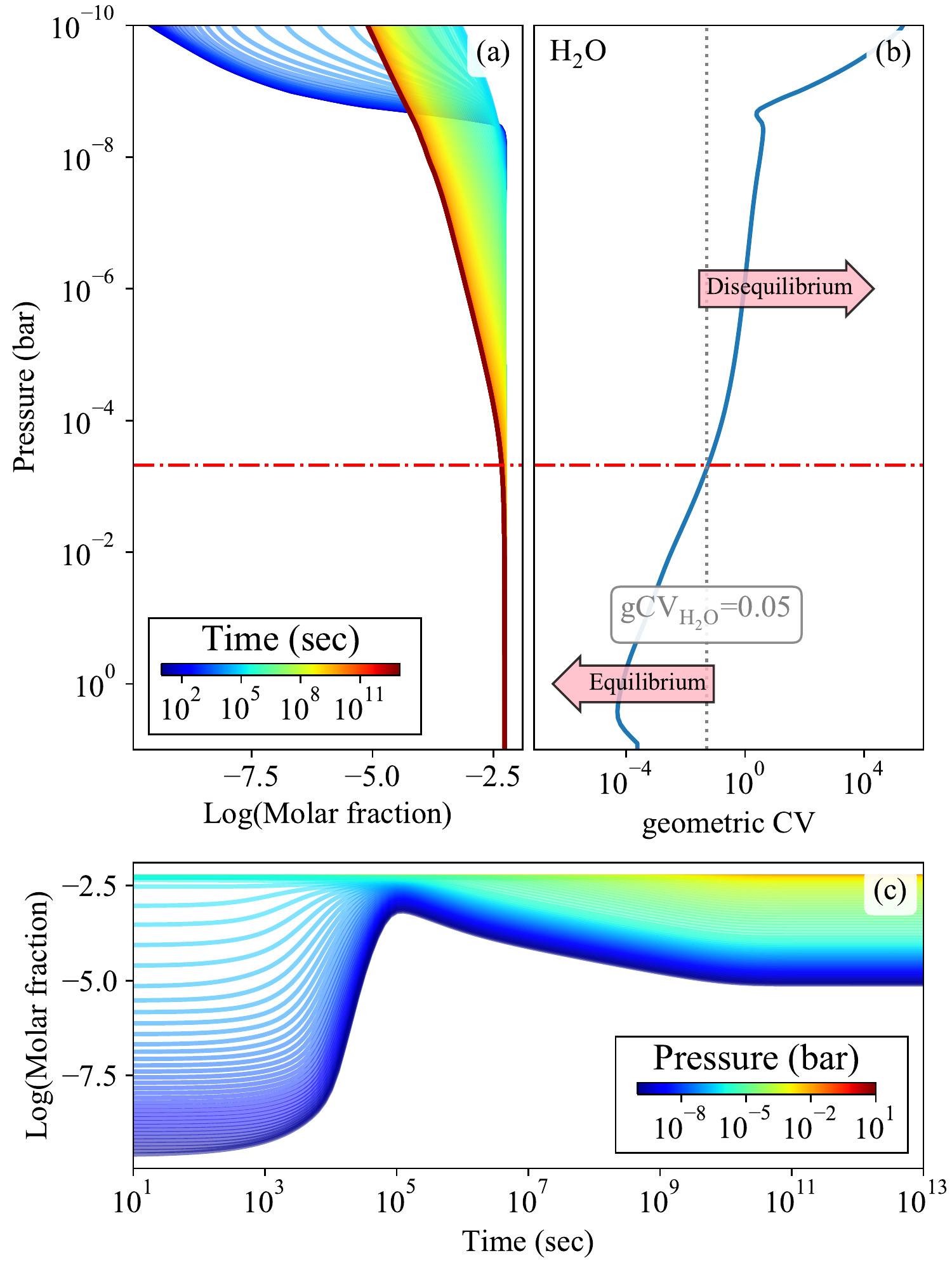}
\caption{A quantitative definition of disequilibrium. We use the results of our diffusion verification model (presented in \hyperref[fig:reversreac1Ddiff]{Figure}\;\ref{fig:reversreac1Ddiff}) as an example. {\bf (a)} Temporal variation of \ce{H2O} abundance from thermochemical equilibrium (blue) to diffusion equilibrium (red). {\bf (b)} Water remains in thermochemical equilibrium when gCV\textsubscript{\ce{H2O}}$<$0.05 (below the red dash-dotted line at $\sim$0.5\;mbar) and is driven away from the thermochemical equilibrium when gCV\textsubscript{\ce{H2O}}$\geq$0.05  (above $\sim$0.5\;mbar). \revision{The red dash-dotted line marks the quenching point.} {\bf (c)} Temporal evolution of water abundances at different pressures. Invariant abundances at $t>$10$^{11}$\;sec insures diffusion equilibrium.} \label{fig:gCVexample}
\end{figure}

This parameter is a standardized measure of dispersion of a quantity \revision{and has been used in other fields, e.g. in the clinical research \citep{schiff_head--head_2014}, to estimate deviation of particular quantities from its mean value. In this work, we introduce this parameter as a mean to quantitatively estimate the quenching levels.} In the chemical kinetic simulations, this parameter can be calculated as the ratio of the temporal standard deviation of abundances, $s$, to the mean value of abundances for a given species at any given pressure level. In our simulations, we use exponential time steps to calculate abundances, hence a better parameter would be the geometric Coefficient of variation (gCV) of abundances. This can be calculated as follows:

\begin{equation}
{\displaystyle {\rm gCV}={\sqrt {\mathrm {e} ^{s_{\rm {ln}}^{2}}-1}}},
\label{eq:gCV}
\end{equation}

where ${s_{\rm ln}}$ is the sample standard deviation of abundances after a natural log transformation and can be estimated as $s_{\rm ln} = s \ln(10)$. Stronger variation of abundance at any given altitude results in a higher gCV. We find that gCV$_i=$0.05 represents the onset of disequilibrium chemistry of species $i$ very well, hence gCV$_i\geq$0.05 can be used to mark the regions at which abundances have deviated from their thermochemical equilibrium values significantly. An example is given in \hyperref[fig:gCVexample]{Figure}\;\ref{fig:gCVexample} based on the results of our diffusion verification model (presented in \hyperref[fig:reversreac1Ddiff]{Figure}\;\ref{fig:reversreac1Ddiff}), where water remains in thermochemical equilibrium when gCV\textsubscript{\ce{H2O}}$<$0.05 (below the red dash-dotted line at $\sim$0.5\;mbar) and is driven away from it when gCV\textsubscript{\ce{H2O}}$\geq$0.05  (above $\sim$0.5\;mbar). \revision{The bottom panel in \hyperref[fig:gCVexample]{Figure}\;\ref{fig:gCVexample} shows the temporal evolution of abundances at different pressures. At the times longer than $t>$10$^{11}$\;sec, abundances reach to steady state at all pressures.}

%Similarly an atmosphere is at disequilibrium if the geometric mean of gCV for all species, $\widehat{\rm gCV}_{\rm all}$, is higher than 0.05. $\widehat{\rm gCV}_{\rm all}$ is defined as follow:

%\begin{equation}
%{\displaystyle \widehat{\rm gCV}_{\rm all} = \left(\prod _{i=1}^{n} \rm gCV_{i}\right)^{\frac {1}{n}}}.
%\label{eq:gCVall}
%\end{equation}

%%%%%%%%%%%%%%%%%%%%%
%%%%%%%%%%%%%%%%%%%%%

\revisiontwo{We present the results based on the "quench point from each species", gCV$_i$. An alternative method would be to calculate all the timescales relevant for a species at any given pressure. While the alternative method is mathematically correct, our method provides a much simpler picture of the quenching levels based on the actual abundance variation due to the kinetics. Another advantage of our method is that there is no need to trace all the reactions and timescales when calculating the abundances, and the quenching can easily be calculated once the simulations are over. The gCV can be also used in a broader sense to study any abundance deviation from any initial condition as a standard mathematical method for such analysis.}

\begin{figure*}
\includegraphics[width=\textwidth]{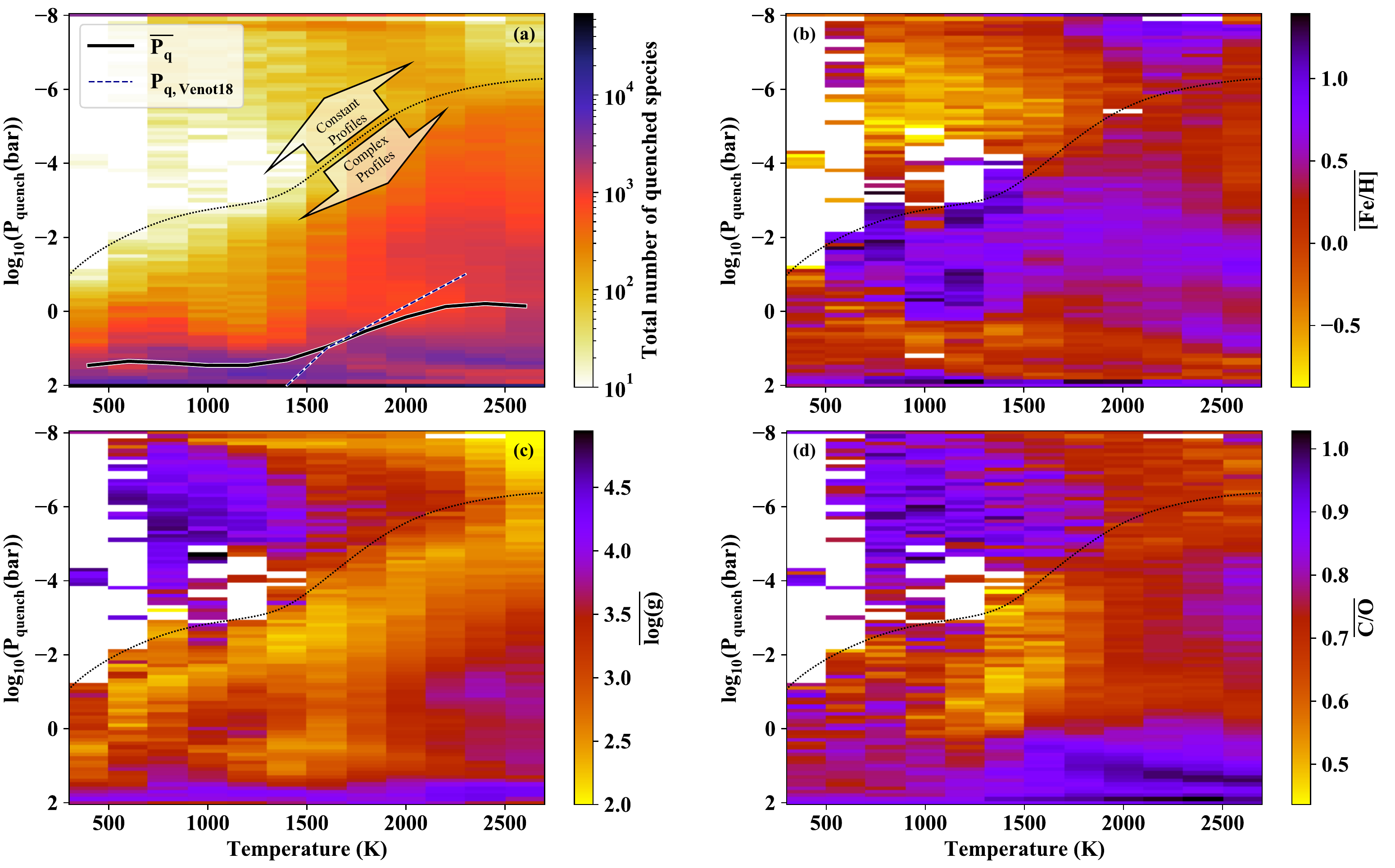}
\caption{A quantitative determination of quenching levels in our grid of chemical kinetic models for K\textsubscript{zz}=10$^{12}$\;cm$^2$ s$^{-1}$ cases. {\bf (a)} Density plot of quenching levels of all species. In general, P\textsubscript{quench} decreases to lower pressures as temperature increases, but its variation at any given temperature is large. However, the average P\textsubscript{quench} (solid black line) is similar to the \citet{venot_better_2018} results of P\textsubscript{quench} for K\textsubscript{zz}=10$^{12}$\;cm$^2$ s$^{-1}$, dashed line. Constant abundance profiles of abundant species cause estimation of P\textsubscript{quench} at lower pressures (above dotted line). {\bf (b,c,d)} Maps of average [Fe/H], log(g), and C/O to graphically demonstrate P\textsubscript{quench} degree of dependence to these parameters.} \label{fig:pqall}
\end{figure*}

By using this new metric we investigate the quenching levels in our grid of models. \hyperref[fig:pqall]{Figure}\;\ref{fig:pqall} (a) shows the density plot of quenching levels of all species in our K\textsubscript{zz}=10$^{12}$\;cm$^2$ s$^{-1}$ models. In general, P\textsubscript{quench} decreases to lower pressures as temperature increases, but its variation at any given temperature is large. For instance, models with T\textsubscript{eff}$=$2600\;K have shown quenching at all pressures ranging from 100 to 10$^{-8}$bar in these simulations. But, if we calculate the average P\textsubscript{quench} along T\textsubscript{eff} (solid black line) the result is similar to the \citet{venot_better_2018} results of P\textsubscript{quench} for K\textsubscript{zz}=10$^{12}$\;cm$^2$ s$^{-1}$, dashed line in \hyperref[fig:pqall]{Figure}\;\ref{fig:pqall} (a).

We find that for a given mixing strength, quenching levels strongly depend on the effective temperature of the planet, but it also depends on the complexity of abundance profiles. For instance, in most cases if a species is the dominant species, then its abundance profile tends to be almost constant. This constant shape of the profiles leads to an underestimation of their P\textsubscript{quench} (their abundance does not change very much by mixing since it is constant already, therefore gCV$_i$ remains small at higher pressures and a lower P\textsubscript{quench} will be determined). This ``constant profiles'' region is marked by a dotted line in \hyperref[fig:pqall]{Figure}\;\ref{fig:pqall}.

The role of T\textsubscript{eff} (and K\textsubscript{zz}) on the determination of P\textsubscript{quench} is rather obvious by following the chemical versus mixing timescale argument; e.g. see \hyperref[subsec:eddy]{Section}\;\ref{subsec:eddy}. However, relating atmospheric parameters to the complexity of abundance profiles is not trivial. Nevertheless, we find that all atmospheric parameters that we have investigated (i.e. T\textsubscript{eff}, log(g), [Fe/H], and C/O) affect the complexity of abundance profiles up to some degree. The averaged value of [Fe/H], log(g), and C/O (calculated in each 2D histogram bin represented in the panel (a) of \hyperref[fig:pqall]{Figure}\;\ref{fig:pqall}) are shown in \hyperref[fig:pqall]{Figure}\;\ref{fig:pqall} panels (b), (c), and (d) respectively, to present a graphical demonstration of these dependencies. \revisiontwo{While \citet{venot_better_2018} also explored the effect of C/O ratio on the quenching levels, they only found negligible variation. This is mainly due to their narrower parameter-space coverage and their methodology for the determination of the quenching levels as an integrated quantity over the main species.}

We will use gCV$_i$ to present an empirical equation for P\textsubscript{quench}([Fe/H],log(g),C/O) in a follow-up paper.

%%%%%%%%%%%%%%%%%%%%%
\subsection{\revision{Observability of the disequilibrium chemistry with JWST}}  \label{subsec:JWST}

\revision{One of the key upcoming missions to address the diversity and characterization of exoplanetary atmospheres is JWST. Naturally, the first question would be: ``Is there any sweet-spot in the parameter-space to detect the effect of disequilibrium chemistry in the planetary spectra by JWST?''. In order to investigate this, we take a simple approach by subtracting the transmission spectra of each atmospheric model at thermochemical equilibrium and disequilibrium chemistry. The wavelength range spans 0.8\;$\mu$m to 20\;$\mu$m, covering most of the JWST's wavelength range. The shorter wavelengths, i.e. 0.6\;$\mu$m to 0.8\;$\mu$m, would be likely influenced by the Rayleigh scattering and wavelengths longer than 20\;$\mu$m do not show significant molecular features; hence excluded from this analysis. We then find the maximum spectral deviation for each model due to the disequilibrium.}

\revision{We should, however, note that using the measured radius at one wavelength is not adequate for a robust detection of disequilibrium fingerprint on the spectra due to the reference pressure degeneracy. Therefore, this analysis should be taken only as a first step toward a spectrum-sensitivity analysis and not a retrieval-degeneracy analysis.}

\begin{figure*}
\includegraphics[width=\textwidth]{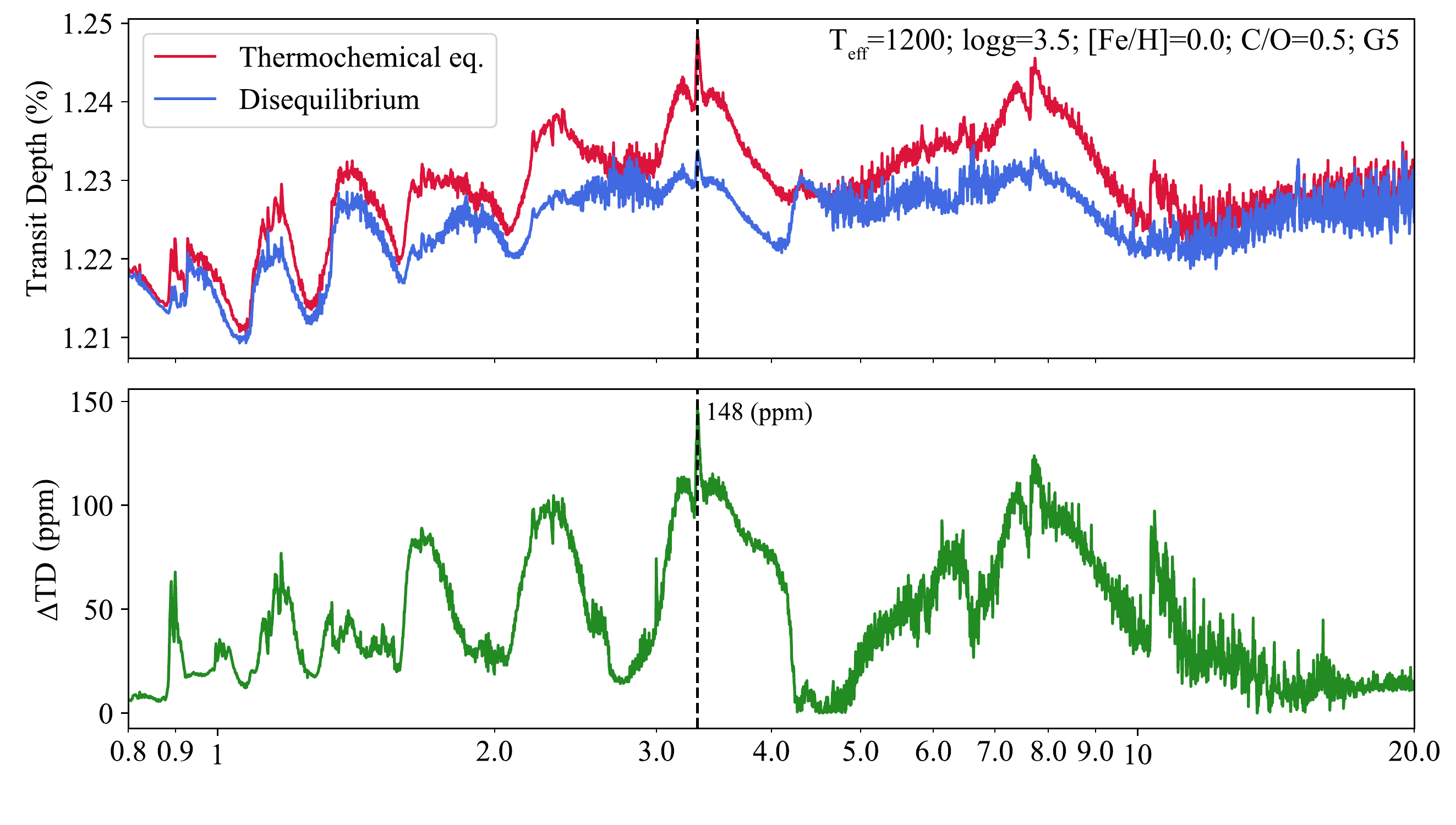}
\caption{\revision{Variation of transmission spectrum due to disequilibrium chemistry for a Jupiter-sized planet orbiting a G5-type star with solar metallicity, C/O$\sim$0.5, and surface gravity of 3.5. {\bf Top)} Methane abundance decreases as a result of added vertical mixing with K\textsubscript{zz}=10$^{12}$\;cm$^2$ s$^{-1}$ (blue) with respect to the thermochemical equilibrium state (red). {\bf Bottom)} The difference between the two spectra indicates that the \ce{CH4} feature at 3.3\;$\mu$m has the maximum variation within JWST's wavelength coverage.}} \label{fig:dTDExample}
\end{figure*}

\begin{figure}
\includegraphics[width=\columnwidth]{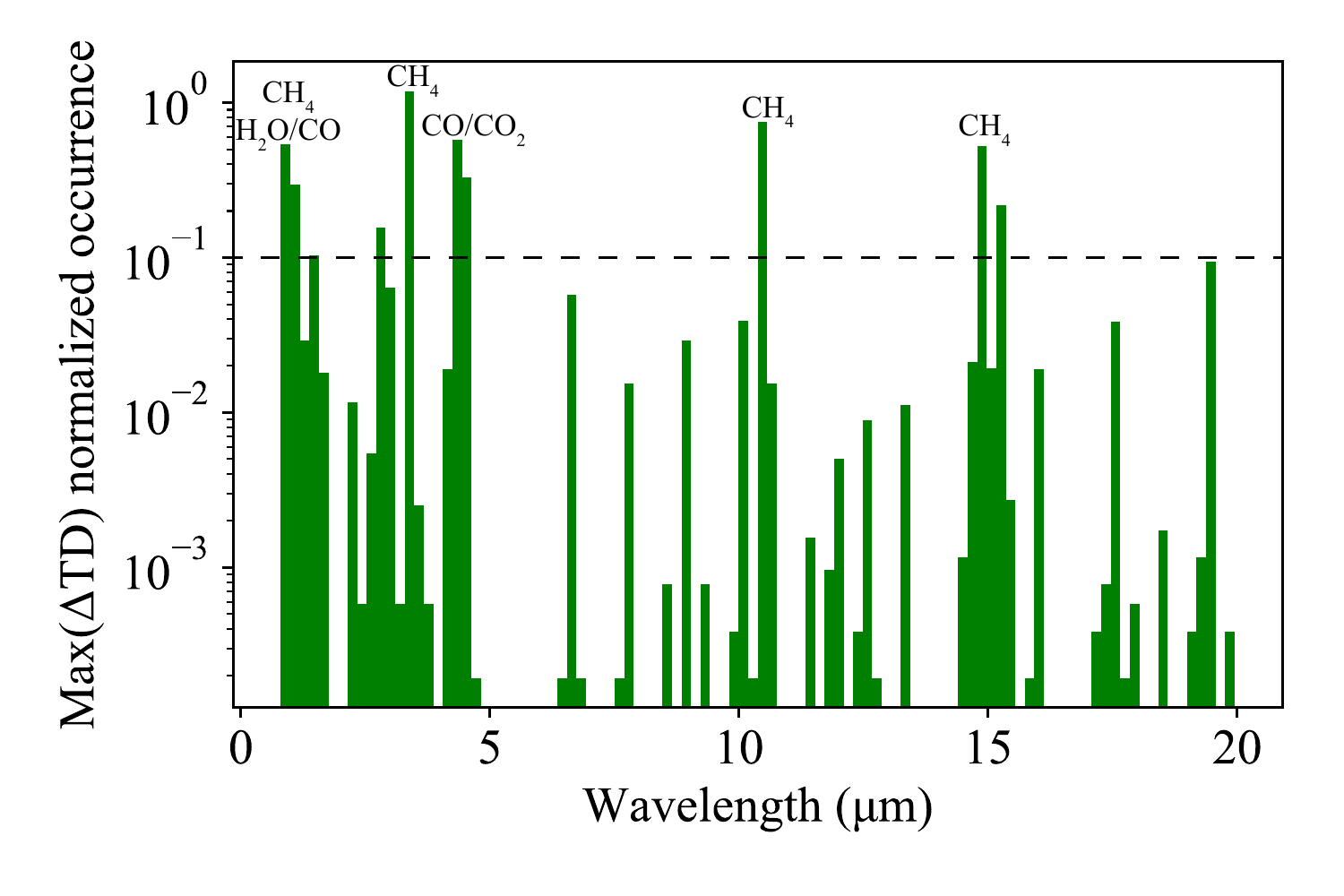}
\caption{\revision{Occurrence rate of maximum variation in the transmission spectra, max($\Delta$TD), within JWST's wavelength range. Five wavelength regions show the highest occurrences, namely $\sim$1\;$\mu$m (\ce{CH4}, \ce{H2O}, or \ce{CO}), $\sim$3.3\;$\mu$m (mostly \ce{CH4}), $\sim$4.5\;$\mu$m (\ce{CO2} or \ce{CO}), $\sim$12\;$\mu$m (\ce{CH4}), and $\sim$15\;$\mu$m (\ce{CH4}).}} \label{fig:dTDwln}
\end{figure}

\revision{\hyperref[fig:dTDExample]{Figure}\;\ref{fig:dTDExample} shows an example of the transmission spectra for a Jupiter-sized planet around a G5-type star, with solar metallicity, C/O$\sim$0.5, and a surface gravity of 3.5 under two conditions: the thermochemical equilibrium and diffusion equilibrium. The diffusion equilibrium model is shown for K\textsubscript{zz}=10$^{12}$\;cm$^2$ s$^{-1}$. In this example, the maximum difference between the spectra of the two models, $\Delta$TD, occurs at the \ce{CH4} feature at $\sim$3.3\;$\mu$m. The difference is about 150 ppm in this model; see the bottom panel of the same figure. While this \ce{CH4} spectral feature is one of the most prominent features to detect the fingerprints of disequilibrium chemistry, other spectral features from \ce{CH4}, \ce{H2O}, \ce{CO2}, or \ce{CO} could be also used as the disequilibrium tracing features. To produce \hyperref[fig:dTDwln]{Figure}\;\ref{fig:dTDwln}, we analyzed where in a spectrum the $\Delta$TD was maximal for all spectra in the grid and calculated the occurrence at different wavelengths. By finding these occurrence rates we determine which wavelengths are likely to have the highest sensitivity to the vertical mixing. We identify five spectral regions with the highest occurrence rates at $\sim$1\;$\mu$m (\ce{CH4}, \ce{H2O}, or \ce{CO}), $\sim$3.3\;$\mu$m (mostly \ce{CH4}), $\sim$4.5\;$\mu$m (\ce{CO2} or \ce{CO}), $\sim$12\;$\mu$m (\ce{CH4}), and $\sim$15\;$\mu$m (\ce{CH4}). Hence, investigation of these wavelengths could provide a higher chance of the detection of disequilibrium spectral signature.}

\revision{\citet{blumenthal_comparison_2018} performed a case study comparing the synthetic emission spectra of HD\;189733b, WASP-80b, and GJ\;436b. They found that the most significant differences in the emission spectra of these planets within the wavelength rage of 4 to 5\;$\mu$m are due to \ce{CO2} and \ce{CO} disequilibrium processes. Although our predictions are based on the synthetic transmission spectra, their results are consistent with our findings for the mentioned wavelength range; see \hyperref[fig:dTDwln]{Figure}\;\ref{fig:dTDwln}.}

\begin{figure*}
\includegraphics[width=\textwidth]{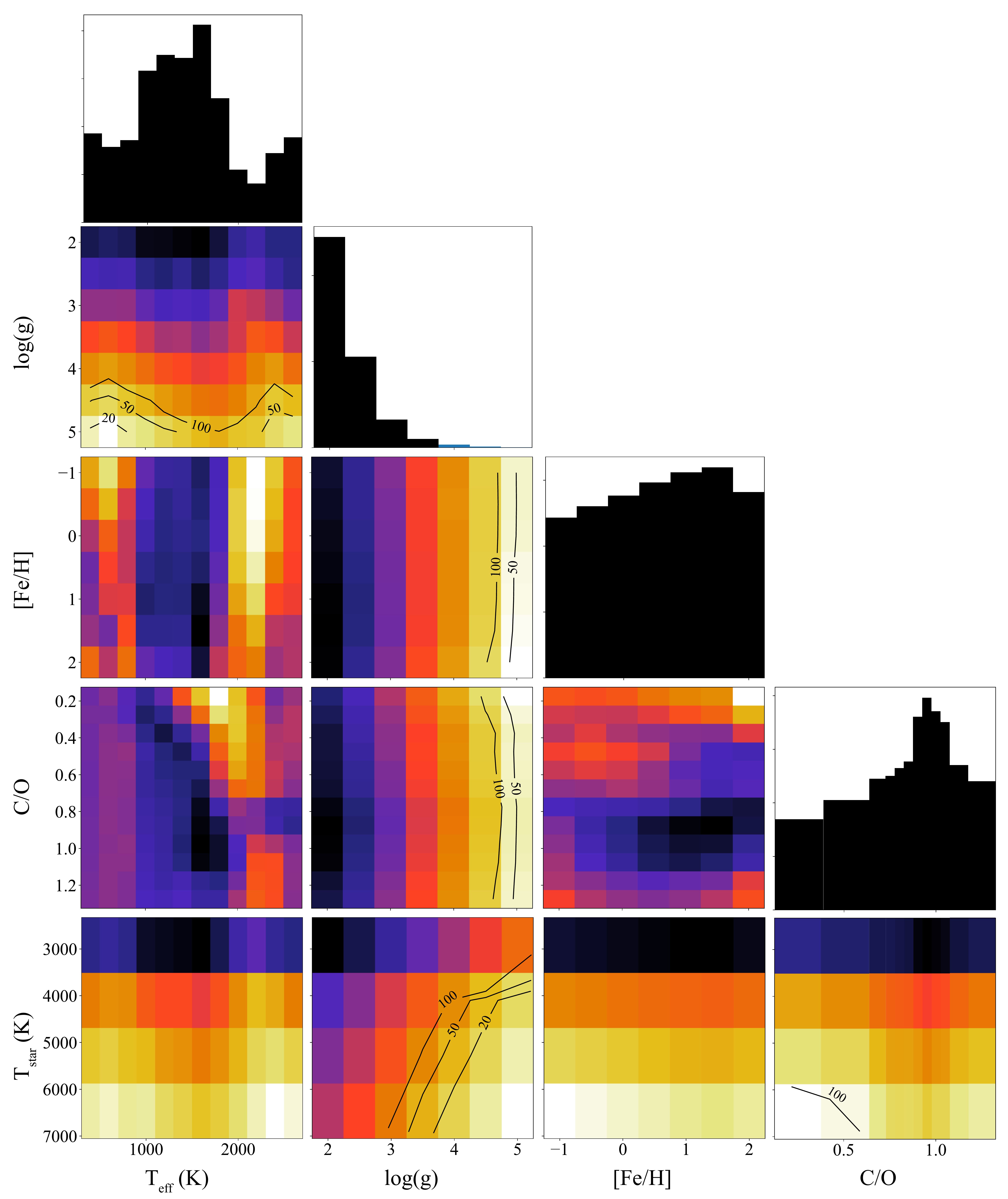}
\caption{\revision{The average maximum variation in the transmission spectra, max($\Delta$TD), due to the disequilibrium for a Jupiter-size planet. Dark-blue represents the parameter-space with the highest max($\Delta$TD). Colormaps are scaled for each panel to show the patterns in more detail. Regions with 20, 50, and 100\;ppm levels are shown with counters; 20 and 50\;ppm represent the noise levels for NIRISS SOSS and MIRI LRS, respectively \citep{greene_characterizing_2016}. Panels without the counters have max($\Delta$TD) higher than 100\;ppm. The marginal panels (1D black histograms at the top) represent $\Delta$TD of individual atmospheric parameters.}} \label{fig:dTDSummary}
\end{figure*}

\revision{By averaging the maximum $\Delta$TD over the free parameters domain, i.e. T\textsubscript{eff}, log(g), [Fe/H], C/O ratio, and the spectral type of the star (or alternatively its surface temperature, i.e. T\textsubscript{star}), we identify the parameter-space at which the transmission spectra are highly sensitive to the vertical mixing. The results are shown in \hyperref[fig:dTDSummary]{Figure}\;\ref{fig:dTDSummary}. The T\textsubscript{eff}-log(g) panel, at the top-left, suggests a strong dependency of disequilibrium detectability on the surface gravity of the planet. While this is not surprising, this panel shows a higher probability of disequilibrium detection at T\textsubscript{eff} between 1000 and 2000\;K. This region coincides with the Class-II and Class-III planets, where the evaporation of condensates plays a crucial role in the photospheric chemistry of planets. In fact, all T\textsubscript{eff}-dependent results in the left column support this finding. Moreover, the T\textsubscript{eff}-C/O panel reveals an interesting trend: as T\textsubscript{eff} increases, a higher chance of disequilibrium detectability occurs at higher C/O ratios. More precisely, there is a region around C/O$\sim$0.95 with a higher probability of disequilibirum detection. This coincides with the minimum-IR opacity/inversion temperature profiles \citep[see e.g.][]{molliere_model_2015}. It appears that such atmospheric condition makes the chemistry to be more easily driven away from its thermochemical equilibrium state. Therefore, this feedback could make the inversion due the minimum-IR opacity, chemically and radiatively unstable. A further self-consistent disequilibirum chemistry calculation must be performed in order to quantitatively assess the significance of this feedback on the inversion.}

\revision{The T\textsubscript{star}-dependent panels, bottom row in \hyperref[fig:dTDSummary]{Figure}\;\ref{fig:dTDSummary}, illustrate a higher probability of disequilibrium detection when the host stars are colder. This is solely due to the fact that colder stars are associated with smaller radius; hence a larger transit depth for a given planetary radius. The T\textsubscript{star}-log(g) panel also suggests that the detection of disequilibrium for a Jupiter-sized planet with high surface gravity maybe only possible for M-dwarfs.}

\revision{A higher metallicity and C/O ratio results in a higher concentration of carbon-bearing species; in particular \ce{CH4}, \ce{CO2}, and \ce{CO}. Hence, a higher detectability of disequilibrium is expected at these conditions. This is shown in the [Fe/H]-C/O panel, where the region of interest (with dark-blue colors) is associated with super-solar metallicities and super-solar C/O ratios.}

\revision{In summary, we find that the detectability of disequilibrium could be maximized by focusing on the targets with T\textsubscript{eff} between 1000 and 1800\;K, orbiting around M-dwarfs, having low surface gravity but high metallicity and a C/O ratio value around unity.}

%%%%%%%%%%%%%%%%%%%%%
%%%%%%%%%%%%%%%%%%%%%

%%%%%%%%%%%%%%%%%%%%%
%%%%%%%%%%%%%%%%%%%%%

%%%%%%%%%%%%%%%%%%%%%
%%%%%%%%%%%%%%%%%%%%%

%%%%%%%%%%%%%%%%%%%%%%%%%%%%%%%%%%%%%%%%%%%%%%%%%%%%%%%%%%%
\subsection{Classification and Color-diagram}  \label{subsec:Classification}

In the first paper in this series \citep[][]{molaverdikhani_cold_2019}, we proposed a new classification scheme for irradiated gaseous planets based on a grid of cloud-free self-consistent models with four classes. Class-I includes planets colder than 600$-$1100\;K. Class-II planets are hotter than Class-I ranging from 600$-$1000\;K to about 1650\;K. The boundary between Class-I and II changes with log(g), metallicity, and stellar type of the host star. Class-III planets have a temperature between 1650\;K and 2200\;K, and Class-IV planets are hotter than 2200\;K. Here we employ the results of our grid of disequilibrium chemistry models to study the effect of vertical mixing on the atmospheric spectra and hence our proposed classification scheme.

\begin{figure*}
\includegraphics[width=\textwidth]{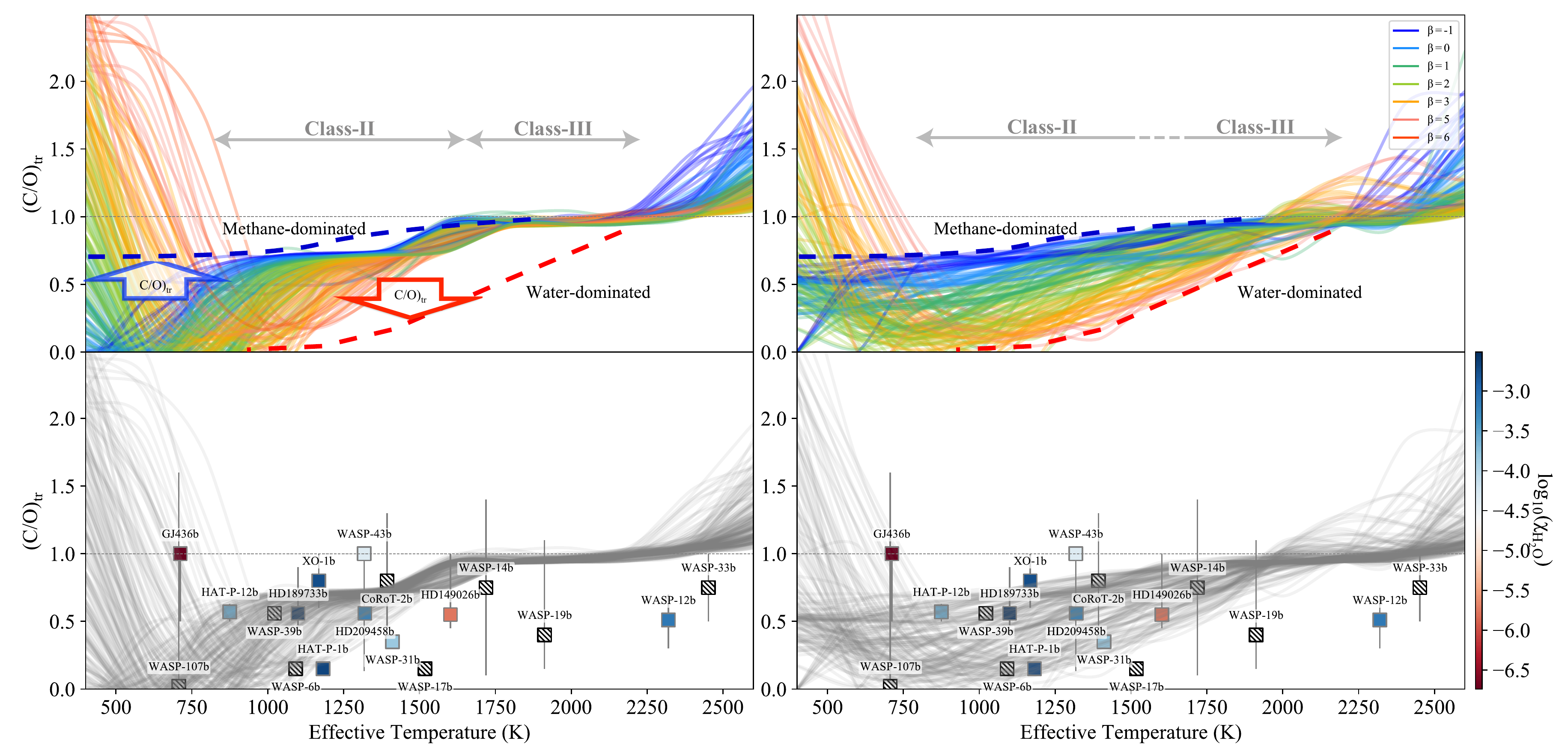}
\caption{The effect of vertical mixing on the transition C/O ratios. {\bf Upper left)} Transition C/O ratios of thermochemically equilibrium models presented in \citet{molaverdikhani_cold_2019}. \revision{The transmission spectra above the transition lines are expected to be \ce{CH4}-dominated and below them to be \ce{H2O}-dominated. Different colors represent the transition C/O ratio at different $\beta$ factors.} {\bf Bottom left)} The location of exoplanets on the thermochemical equilibrium map with estimated T\textsubscript{eff} and C/O. {\bf Upper right)} Transition C/O ratios of chemical kinetic models. {\bf Bottom right)} The location of exoplanets \revision{with estimated C/O ratio,} on the chemical kinetic map. \revision{The lines in the bottom panel are gray to increase the visibility of colors for the water content of the observed planets.}} \label{fig:COtr}
\end{figure*}

As discussed in the first paper, carbon-to-oxygen ratio affects the composition of hot planetary atmospheres in a way that water is expected to be more prominent at lower C/O ratios and methane, CO, or HCN to be the dominant chemical products at higher C/O ratios. The spectral decomposition technique provides the necessary tool to find the transition C/O ratio, i.e. the C/O ratio at which the atmospheric spectra changes from a water- to methane-dominated spectrum. By employing this technique, we estimate the transition C/O ratios for our grid of disequilibrium chemistry models, see  \hyperref[fig:COtr]{Figure}\;\ref{fig:COtr}, for K\textsubscript{zz}=10$^{12}$\;cm$^2$ s$^{-1}$ models. 

%%%%%%%%%%%%%%%%%%%%%
%The vertical mixing makes the transition from methane- to water-dominated atmosphere to occur at lower C/O ratios on average for Class-II and III planets. Higher $\beta$ values (a linear combination of surface gravity and metallicity) represent deeper photospheres, which are more prone to the vertical mixing. Planets at the boundary of Class-II and III are affected by diffusion the most (their C/O\textsubscript{tr} has decreased the most). Vertical mixing seems to merge Class-II and III to one extended class as well, covering the whole range of $\sim$900\;K to $\sim$2000\;K in one class.
%%%%%%%%%%%%%%%%%%%%%

In general, but not always, vertical mixing on Class-II and III planets cause the transition C/O ratios to decrease for atmospheres with high $\beta$. The $\beta$ factor is a linear combination of surface gravity and metallicity. Higher $\beta$ values represent deeper photospheres, i.e. at higher pressures. See \citep[][]{molaverdikhani_cold_2019} for a detailed discussion on the $\beta$ factor. This is mostly caused by the higher amount of \ce{CH4} at the photospheric level, when $\beta$ is higher. Its effect is different for atmospheres with low $\beta$ and vertical mixing increases the transition C/O ratios. These are shown by blue (for low $\beta$ models) and red (for high $\beta$ models) vectors on \hyperref[fig:COtr]{Figure}\;\ref{fig:COtr} left panel.

The only difference between Class-II and Class-III planets is that in Class-II the atmosphere still contains some oxygen-bearing condensates, but in Class-III planets all those condensates are evaporated. This causes the transition C/O ratios in Class-III to become independent of atmospheric conditions and remain at a constant value of $\sim$0.9. Vertical mixing removes this difference and merges Class-II with III to one extended class; covering the whole range of $\sim$900\;K to $\sim$2000\;K in one class.

In Class-I planets, both \ce{H2O} and \ce{CH4} are usually the dominant species and have relatively less vertically variant profiles. Hence, their abundance variation at TOA is less sensitive to the vertical mixing. Class-IV planets with C/O$<$1.0 are mostly deficient of \ce{CH4} due to their hot temperatures. Hence their transition C/O ratio remains higher than unity at all time and are usually insensitive to the vertical mixing. However, some Class-IV planets, T\textsubscript{eff}$>$2000\;K, with very high $\beta$, $>5$, show an increase in their transition C/O ratio due to lower \ce{CH4} abundances at high pressures. Consequently, our proposed four classes of planets based on our grid of cloud-free models can change to a three classification scheme of planets assuming strong vertical mixing in their atmospheres.

Regardless of the naming of these classes, there exists a ``Methane Valley'' in both thermochemical equilibrium and vertical mixing cases (see \hyperref[fig:COtr]{Figure}\;\ref{fig:COtr}). The Methane Valley is a region between 800 and 1500\;K, where methane is expected to cause the dominant spectral features in the transmission spectra of planets. Hence, a greater chance of \ce{CH4} detection is expected for the planets within this region. The first robust \ce{CH4} detection on an irradiated planet was indeed reported for HD\;102195b \citep{guilluy_exoplanet_2019}. This is a Class-II planet with T\textsubscript{eq}$\sim$963\;K, within the Methane Valley and consistent with our prediction.

\begin{figure*}
\includegraphics[width=\textwidth]{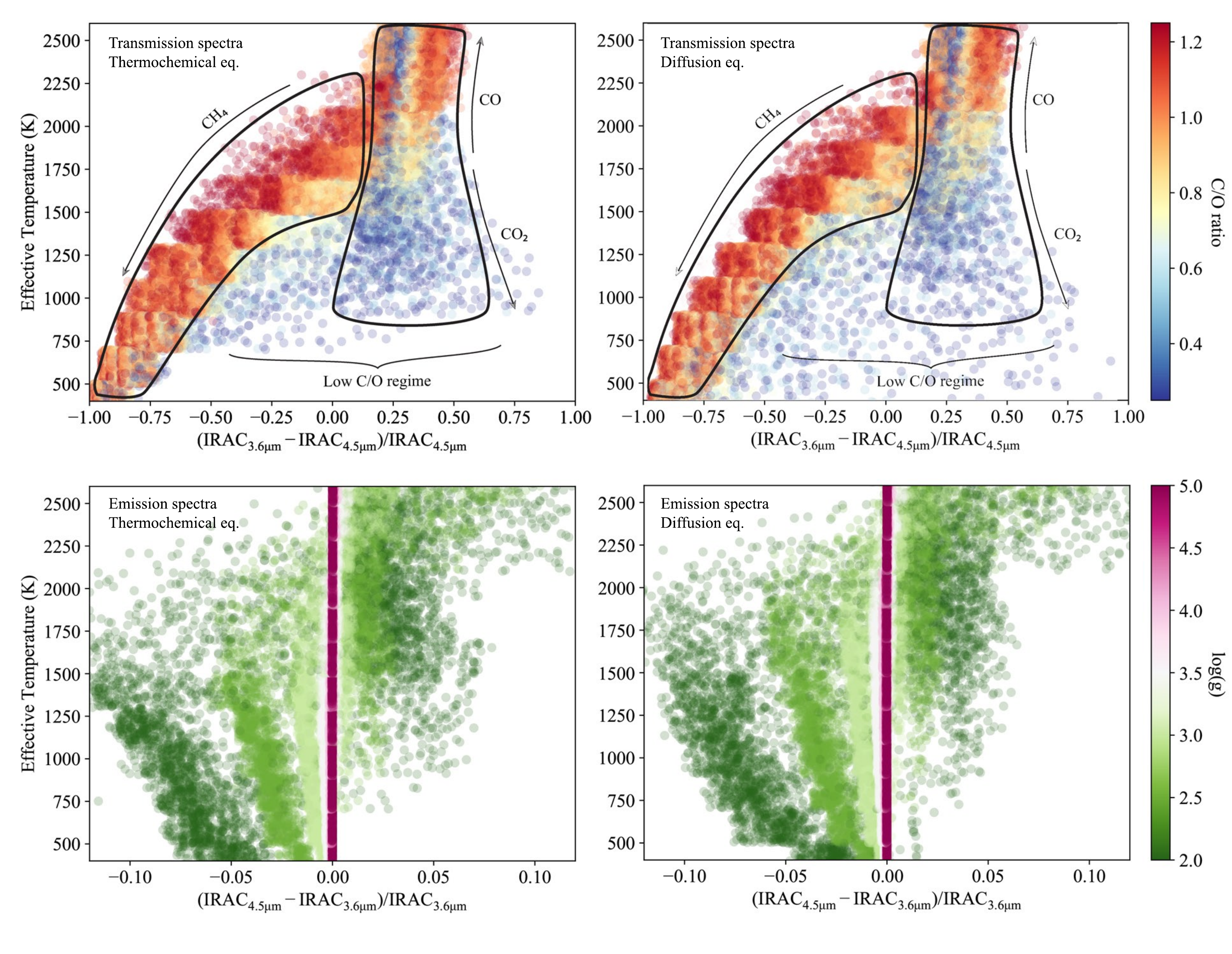}
\caption{Synthetic Spitzer IRAC color–temperature diagrams for cloud-free atmospheres under thermochemical equilibrium ({\bf left panels}) and diffusion equilibrium ({\bf right panels}) conditions. {\bf Top panels} Color diagram based on emission spectroscopy, i.e., IRAC data describes the secondary eclipse depth at $\lambda$\;($\mu$m). {\bf Bottom panels} Same for the transmission technique, i.e., IRAC data describes the transit depth at $\lambda$\;($\mu$m). No significant difference can been seen between thermochemical and diffusion equilibrium models, except for Class-I planets with very low C/O ratios.} \label{fig:colormap}
\end{figure*}

We also calculate the transmission and emission color-maps, using Spitzer IRAC channel 1 and 2, similar to what we presented in the first paper. We note small changes seen in the IRAC emission maps, when going from thermochemical equilibrium to disequilibrium chemistry. Vertical mixing is expected to play a strong role in the atmospheres of cool (T $<$ 1000\;K), self-luminous atmospheres with low (planet-like) log(g) \citep[see, e.g.,][]{zahnle_methane_2014}. In such atmospheres the abundances of the cooler upper atmospheric layers is overruled by the abundances mixed up from hot layers at higher pressures, which leads to a significant increase in CO abundance, at the expense of \ce{CH4}. A similar effect appears to occur for a few hundred atmospheres in our grid of $+$28,000 models, which start to fill in the region with (F$_{3.6}$-F$_{4.5}$)$/$F$_{4.5}$ $>$ -0.5. This happens for the models with effective temperatures below $\sim$900\;K and with very small C/O ratios ($<$0.25) only (blue points in the low C/O regime in \hyperref[fig:colormap]{Figure}\;\ref{fig:colormap}). Considering what has been predicted for self-luminous atmospheres, this is unexpected. A likely reason for this difference is that the atmospheric temperature profiles of irradiated planets are more isothermal than those of self-luminous ones, such that the deeper (higher pressure) regions from which material is mixed up is cooler, and hence richer in \ce{CH4}\footnote{CO to \ce{CH4} conversion is favored at cool, high pressure conditions; \revision{see e.g. \citet{lodders_chemistry_2006,marley_clouds_2002,hubeny_model_2017,noll_onset_2000,skemer_first_2012,saumon_non-equilibrium_2003}}.}. 

The color maps do not show any other statistically meaningful differences, see \hyperref[fig:colormap]{Figure}\;\ref{fig:colormap}, and the two main populations (i.e. \ce{CH4}-driven and CO/\ce{CO2}-driven populations) remain almost invariant. Our results indicate that deviations from the Spitzer equilibrium chemistry color map presented here are most likely due to effects different than the disequilibrium processes studied in our work, with clouds being a strong contender, the effect of which we will study in our upcoming third paper of this series. If disequilibrium should be the culprit for such deviations, it would point to very small C/O ratios ($<$ 0.25).

%%%%%%%%%%%%%%%%%%%%%
%%%%%%%%%%%%%%%%%%%%%

%%%%%%%%%%%%%%%%%%%%%%%%%%%%%%%%%%%%%%%%%%%%%%%%%%%%%%%%%%%
%%%%%%%%%%%%%%%%%%%%%%%%%%%%%%%%%%%%%%%%%%%%%%%%%%%%%%%%%%%
%%%%%%%%%%%%%%%%%%%%%%%%%%%%%%%%%%%%%%%%%%%%%%%%%%%%%%%%%%%
\section{Conclusion} \label{sec:discussion}

In this paper, we introduced our newly developed Chemical Kinetic Model (\texttt{ChemKM}). The code includes a variety of atmospheric processes including photolysis, molecular and eddy diffusion, and condensation and rainout. Other processes, such as atmospheric ablation and escape can also be included. It is also possible to include galactic and solar cosmic rays (GCRs) and scattered UV sources by LIPM in the models.

After verification of individual processes in \texttt{ChemKM}, we compared our results to the \citet{venot_photochimie_2012} model for HD\;189733b. The calculated abundances were consistent between the models except at the $\mu$bar regime, where the molecular diffusion and photochemistry are the dominant processes. As a conclusion, we recommend careful consideration of these processes when abundances at TOA have to be calculated.

We used our grid of disequilibrium chemistry atmospheres \revision{with six free parameters, planetary effective temperature (T\textsubscript{eff}), surface gravity (log(g)), metallicity ([Fe/H]), carbon-to-oxygen ratio (C/O), spectral type of the host star, and vertical mixing (K\textsubscript{zz}=10$^6$, 10$^9$, and 10$^{12}$\;cm$^2$ s$^{-1}$),} to determine the quenching levels. We propose a new metric, the geometric Coefficient of variation (gCV) of abundances, for a quantitative measure of quenching. We find P\textsubscript{quench} varies significantly under our self-consistent static TP structure setup. We find that all atmospheric parameters ([Fe/H], log(g), and C/O) affect P\textsubscript{quench} by changing the atmospheric composition.

\revision{To explore the detectability of disequilibrium spectral fingerprints due to the molecular and eddy diffusion by JWST, we recommend focusing on the targets with T\textsubscript{eff} between 1000 and 1800\;K, orbiting around M-dwarfs, having low surface gravity but high metallicity and high C/O (see \hyperref[subsec:JWST]{Section}\;\ref{subsec:JWST}).} 

We also find that the ``Methane Valley'' remains largely unchanged by the inclusion of vertical mixing. This is a region between 800 and 1500\;K, where a greater chance of \ce{CH4} detection is expected. Indeed, the first robust \ce{CH4} detection on an irradiated planet was within this region \citep{guilluy_exoplanet_2019}; supporting our prediction from \citet{molaverdikhani_cold_2019}. \revision{We argued in Paper-I that the detection of \ce{CH4} or the lack of such detection, could hint at the prevalence of cloud formation or disequilibrium chemistry. Here we find that disequilibrium unlikely to change this picture significantly. Thus, characterization of planetary atmospheres within this parameter space is expected to provide a diagnostic tool to identify cloud formation condition.}

Our further analysis showed that the two main populations in the Spitzer IRAC's emission color-maps remain largely unchanged, when including mixing. Although we note some differences between the two maps for the models with effective temperatures below $\sim$900\;K and with very small C/O ratios ($<$0.25), see \hyperref[fig:colormap]{Figure}\;\ref{fig:colormap}. Therefore any deviation of observational points from these color-maps is likely to be due to the presence of clouds and not disequilibrium chemistry, but the effect of mixing for cold planets with very low C/O ratios could be significant.

Clouds can potentially change the TP structure, composition, and ultimately the spectra of planetary atmospheres. In the next paper in this series, ``From cold to hot irradiated gaseous exoplanets'', we investigate how cloud formation affect the atmospheric spectra and what types of planets are most affected by this process.

\section{Acknowledgment} \label{sec:acknowledgment}
The authors would like to thank Dr. Dmitry A. Semenov for constructive feedbacks during the development of \texttt{ChemKM} and would like to thank Dr. Olivia Venot, Dr. Juliane Moses, Dr. Eric Hébrard, and Dr. Shang-Min Tsai for providing their chemical networks. We used the online database KIDA for some kinetic data. This research has made use of the NASA Exoplanet Archive, which is operated by the California Institute of Technology, under contract with the National Aeronautics and Space Administration under the Exoplanet Exploration Program. \revision{The authors would like to thank the anonymous reviewer for her/his helpful and constructive comments that greatly contributed to improving the final version of the paper.}

%%%%%%%%%%%%%%%%%%%%%%%%%%%%%%%%%%%%%%%%%%%%%%%%%%%%%%%%%%%
%%%%%%%%%%%%%%%%%%%%%%%%%%%%%%%%%%%%%%%%%%%%%%%%%%%%%%%%%%%
%%%%%%%%%%%%%%%%%%%%%%%%%%%%%%%%%%%%%%%%%%%%%%%%%%%%%%%%%%%
\revisiontwo{\section{Softwares} \label{sec:software}
Employed softwares and packages in this work are as follow: \software{petitCODE \citep{molliere_model_2015,molliere_observing_2017}, ODEPACK \citep{hindmarsh_odepack_1983}, and ChemKM (this work).}}

%%%%%%%%%%%%%%%%%%%%%%%%%%%%%%%%%%%%%%%%%%%%%%%%%%%%%%%%%%%
%%%%%%%%%%%%%%%%%%%%%%%%%%%%%%%%%%%%%%%%%%%%%%%%%%%%%%%%%%%
%%%%%%%%%%%%%%%%%%%%%%%%%%%%%%%%%%%%%%%%%%%%%%%%%%%%%%%%%%%
\appendix

%%%%%%%%%%%%%%%%%%%%%%%%%%%%%%%%%%%%%%%%%%%%%%%%%%%%%%%%%%%
\section{Methodology}\label{subsec:methodology}
\subsection{Chemical solvers}\label{subsec:solver}

%%%%%%%%%%%%%%%%%%%%% rtols
\begin{figure*}
\includegraphics[width=\textwidth]{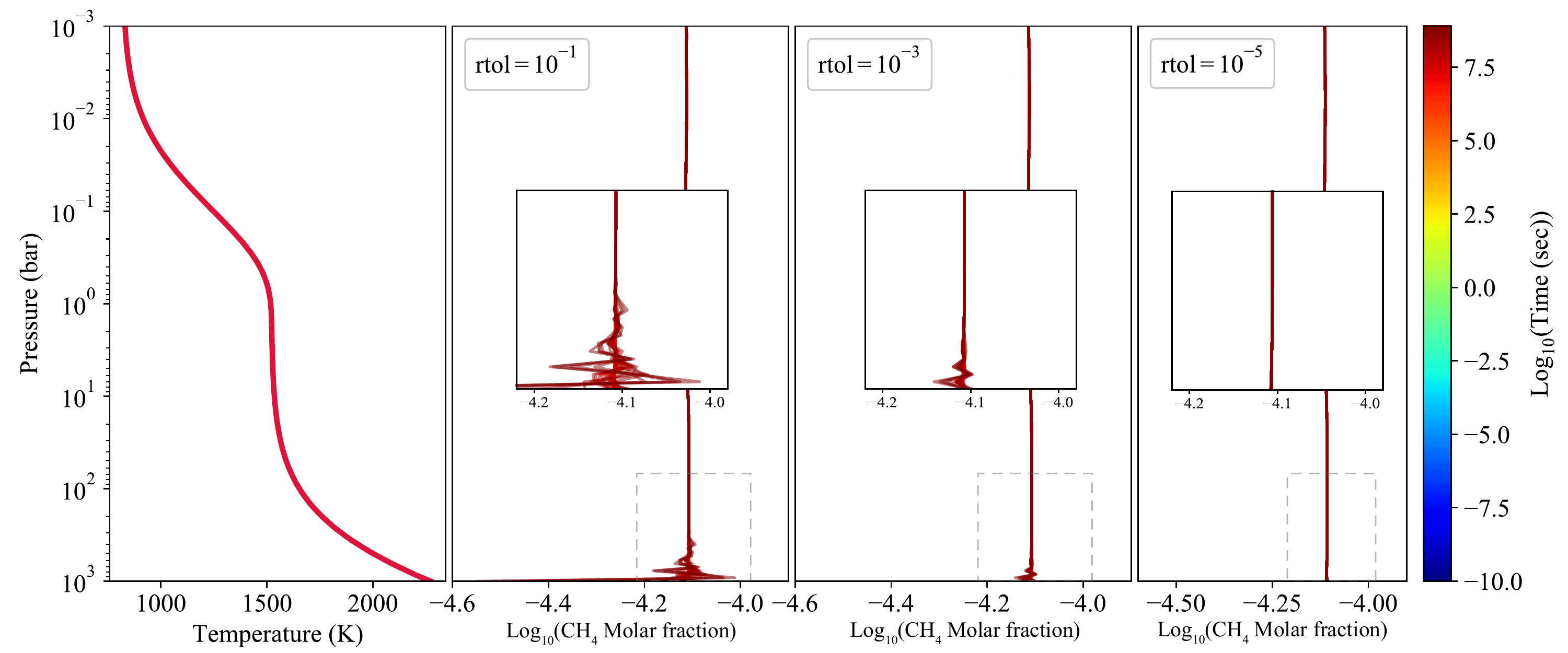}
\caption{Numerical instability due to the choice of relative tolerance, rtol, in the numerical solver. Poor numerical convergence occurs at high temperature-pressure regions of the atmosphere, unless a proper rtol value is chosen.  \textbf{Left)} Temperature profile of a typical Hot Jupiter with T\textsubscript{eff}$=$1500\;K. \textbf{Rest of panels)} show the time evolution of \ce{CH4} vertical abundance. Initial abundances are calculated by Gibbs free minimization and are expected to remain the same throughout the simulation. Three simulations are the same, except their choice of rtol. Simulations with rtol$=10^{-1}$ and $10^{-3}$ show non-negligible numerical instabilities, whereas rtol$=10^{-5}$ results in an adequately accurate abundance estimation. \label{fig:rtols}}
\end{figure*}

The challenge of most chemical kinetic models is to find an efficient method to solve the system of ODEs, which are stiff and sparse in nature. Diversity of physical and computational complexities in the simulation of planetary atmospheres results in different degrees of stiffness, and hence complicates the choice of a fast and yet accurate sparse-stiff ODE solver. We have examined two well-established ODE solver packages from the ODEPACK collection, namely DLSODE (Double precision Livermore Solver for Ordinary Differential Equations) and DVODPK (Double precision Variable-coefficient method for Ordinary Differential equations using Preconditioned Krylov method) \citep{hindmarsh_odepack_1983}.

The DLSODE solver is based on the older GEAR and GEARB packages, and has the same numerical core. In GEAR, the error is controlled only by one scalar relative error tolerance, called EPS. In DLSODE, on the other hand, the user can set both relative and absolute error tolerances, rtol and atol, which optionally can be scalar or vector. The choice of these error tolerances are usually a compromise between the accuracy and speed of the simulation, i.e. a smaller error tolerance makes the solutions more accurate, but at the same time it increases the computational time, because the solver requires more iterations to achieve the specified tolerance. Our extensive examination of rtol and atol indicates that the computational time scales with rtol exponentially, but the choice of atol has no significant effect on the computational time. Therefore, we choose a very small atol, $10^{-99}$, as the default atol value for our simulations.

A choice of rtol$=10^{-1}$, $10^{-2}$ or $10^{-3}$ can be used for a rapid verification of the system's behavior, but we recommend rtol$=10^{-5}$ to insure numerical stability when the production and loss terms are large, but of the same order, i.e. the system is at or close to $\sim$steady state, see \hyperref[fig:rtols]{Figure}\;\ref{fig:rtols}. In this figure we present a worst case scenario where the atmosphere is initialized at its equilibrium state and is forced to remain at that condition by using chemical kinetic calculations. This is not the usual setup and planetary models normally include at least one disequilibrium process such as diffusion. The numerical instability at rtol$=10^{-1}$ and $10^{-3}$ are shown in \hyperref[fig:rtols]{Figure}\;\ref{fig:rtols} are for \ce{CH4} abundance profiles but almost all other species behave similarly. Diffusion usually relaxes the limit on rtol for achieving the numerical stability and introduces a damping effect. In addition, stronger diffusion forces the system to reach the steady state faster; leading to less propagated numerical errors in the steady state solution of the system. Hence, a lower rtol can be selected when stronger diffusion mechanism are in play.

It is also possible to specify the tolerances as vectors; i.e. assigning each species its own relative and absolute tolerances. This treatment could be used, for instance, on a reactant that is completely depleted in at least one vertical layer. In this paper, we use scalar values for rtol and atol with the values of $10^{-5}$ and $10^{-99}$, respectively.

DVODPK employs iterative methods to solve linear algebraic systems. It is based on orthogonal projection techniques, which approximate a solution of the linear system from a Krylov subspace. This technique is more advanced and more efficient in comparison with DLSODE's method \citep{nejad_comparison_2005}, however, its usage requires decomposition and initialization of the Jacobian matrix. This is not a trivial task and our investigations revealed that this technique requires adaptation of the Jacobian decomposition for every new simulation. Moreover, DVODPK's performance changes significantly with the physical conditions considered in the model, such as inclusion of condensation, ablation, outgassing, etc.

\begin{figure}[t]
\includegraphics[width=\columnwidth]{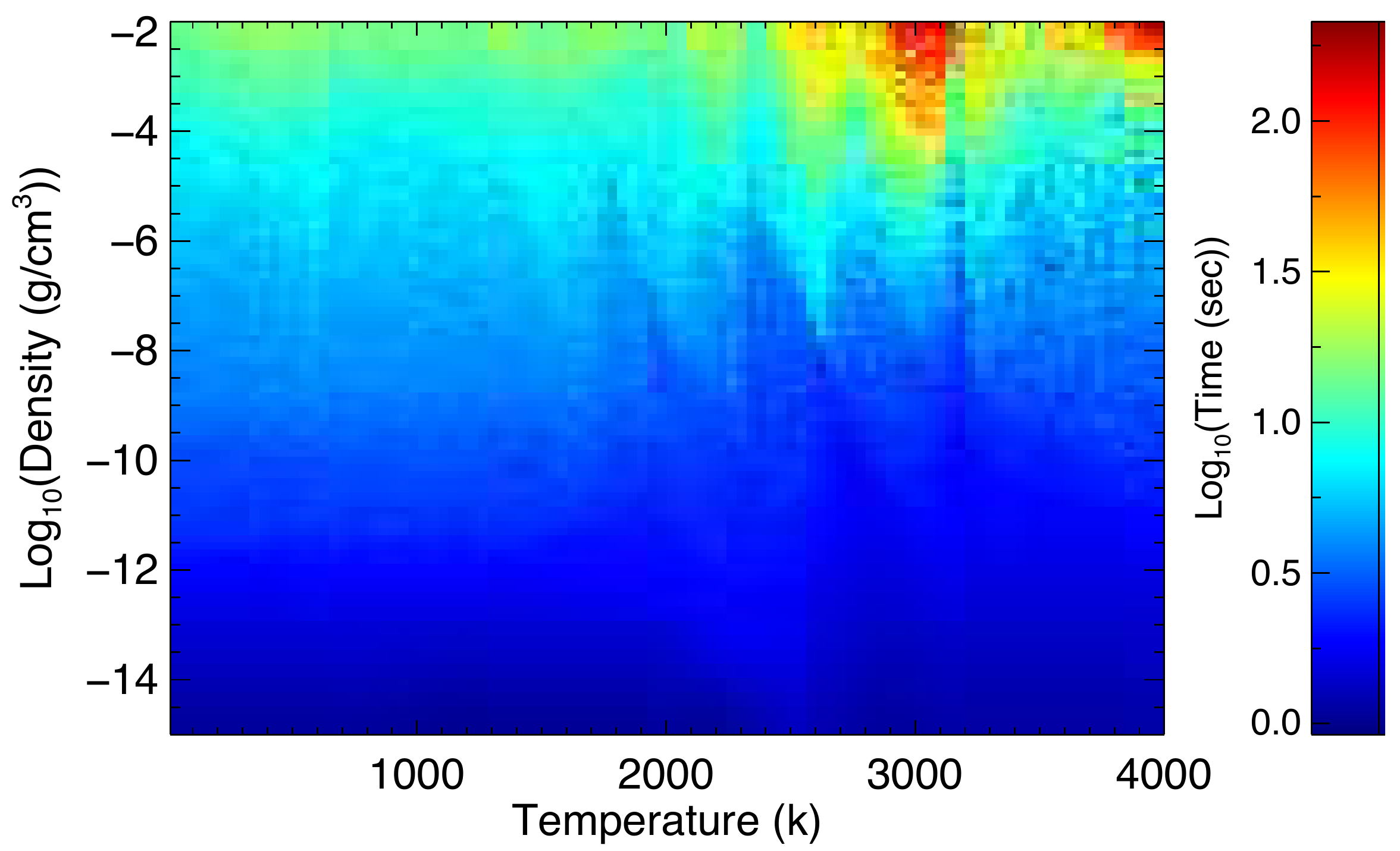}
\caption{DVODPK's running time for a suite of 0D simulations over a broad range of densities and temperatures. The solution of the system of ODEs by DVODPK appears to be less efficient at higher density and temperature conditions. \label{fig:dvopk}}
\end{figure}

On top of all is DVODPK's inefficiency under planetary atmosphere conditions. \hyperref[fig:dvopk]{Figure}\;\ref{fig:dvopk} illustrates the DVODPK's running time for a suite of 0D simulations over a broad range of densities and temperatures. We found DVODPK is less efficient to solve ODEs at higher density and temperature conditions, and hence is not suitable for the application to planetary atmosphere kinetic simulations. Therefore we use DLSODE from the ODEPACK library as our default solver.

%%%%%%%% dvopk running time

%%%%%%%%%%%%%%%%%%%%%%%%%%%%%%%%%%%%%%%%%%%%%%%%%%%%%%%%%%%
\subsection{Chemical networks}\label{subsec:network}
Answering the question of which set of reactions, i.e. chemical network, must be used in a chemical kinetic model is arguably the most important question that should be addressed before setting up a model. One could establish a chemical network from the scratch by doing a standard literature search for the thermodynamic polynomial coefficients, reaction rate constants, UV cross sections, and branching ratios, or alternatively employ the commonly used networks in the field. Here we took the latter approach as the establishment of a new network is beyond the scope of this study. Currently, nine chemical networks are available in \texttt{ChemKM}: Moses's Hot Jupiter network \citep{moses_disequilibrium_2011}, Venot's 2012 network \citep{venot_chemical_2012}, Hebrard's Titan network \citep{hebrard_neutral_2012}, Hebrard's C$_3$H$_p$ network \citep{hebrard_photochemistry_2013}, Venot's full network \citep{venot_new_2015}, Vulcan H-C-O network \citep{tsai_vulcan:_2017}, Moses's ice-giants network \citep{moses_seasonal_2018}, Venot's reduced network \citep{venot_reduced_2019}, and Pearce's HCN network \citep{pearce_consistent_2019}, which differ in the complexity, size (number of species and reactions), type of chemistry (reversible/irreversible, H-C-O species, H-C-O-N species, HCN network), and the temperature range over which the chemical schemes are valid, see \hyperref[tab:tab1]{Table}\;\ref{tab:tab1}.

%%%%%%%%%%%%%%%%%%%
\begin{table*} \centering
%\begin{table}[]
\begin{tabular}{p{3cm}p{1.5cm}p{1.2cm}p{1.4cm}p{3.3cm}}
{\bf Chemical Network}   & {\bf Elements} & {\bf Species} & {\bf Reactions} & {\bf Reference}                                              \\ \hline \hline
Moses Hot Jupiters & H-C-O-N  & 90      & 800      & \citealp{moses_disequilibrium_2011} \\ \hline
Venot 2012         & H-C-O-N  & 103     & 963      & \citealp{venot_chemical_2012}       \\ \hline
Hébrard 2012       & H-C-O-N  & 135     & 788       & \citealp{hebrard_neutral_2012}      \\ \hline
Hébrard 2013       & H-C-O-N  & 90     & 941       & \citealp{hebrard_photochemistry_2013}      \\ \hline
Venot 2015         & H-C-O-N  & 238     & 2011      & \citealp{venot_new_2015}       \\ \hline
VULCAN H-C-O       & H-C-O    & 29      & 300       & \citealp{tsai_vulcan:_2017}         \\ \hline
Moses Ice-giants   & H-C-O    & 69      & 385       & \citealp{moses_seasonal_2018}       \\ \hline
Venot reduced      & H-C-O-N  & 29      & 181       & \citealp{venot_reduced_2019}        \\ \hline
Pearce HCN         & H-C-N    & 11      & 42        & \citealp{pearce_consistent_2019}    \\ \hline

\end{tabular} \caption{General characteristics of chemical networks implemented in \texttt{ChemKM}. He and Ar are excluded from the list of species. The reverse reactions calculated by using the thermodynamic principle of microscopic reversibility are not included.} \label{tab:tab1} \end{table*}
%%%%%%%%%%%%%%%%%%%

\texttt{ChemKM} is able to handle a variety of reaction types including KIDA's standard formulations\footnote{http://kida.obs.u-bordeaux1.fr} (e.g. Modified Arrhenius reactions, third body assisted reactions with Troe \citep{troe_theory_1983,gilbert_theory_1983}, and SRI \citep{stewart_pressure_1989,kee_chemkin-ii_1989} falloff functions) as well as Vuitton \citep{vuitton_rapid_2011} and Jasper \citep{jasper_kinetics_2007} formulations. Therefore \texttt{ChemKM} can easily accept new networks as the input of models.

%%%%%%%%%%%%%%%%%%%%%%%%%%%%%%%%%%%%%%%%%%%%%%%%%%%%%%%%%%%
\subsection{Condensation and rainout} \label{subsec:condrainout}

Observations of planets suggest the presence of clouds nearly in all classes of planets with substantial atmosphere. For instance, Venus is fully covered by a thick cloud layer, containing mostly \ce{H2SO4} particles \citep[e.g.,][]{molaverdikhani_abundance_2012}. On Earth, water clouds produce a patchy scenery, covering about $70\%$ of the surface with optical depth of larger than 0.1 \citep[e.g.,][]{stubenrauch_assessment_2013}. Despite its tenuous atmosphere, conditions on Mars allow for the formation of what is mostly believed to be high-altitude \ce{CO2} clouds \citep[e.g.,][]{clancy_co2_1998}.

Outer planets in the solar system have even colder atmospheres, resulting in more efficient cloud-forming environments. Observations and simulations support the formation of \ce{NH3}, \ce{NH4SH}, and \ce{H2O} Jovian clouds above 10\;bar. Their altitude and thickness, however, could vary both spatially and temporally \citep{pater_peering_2016}. Clouds with similar compositions are expected to form on Saturn; owing to their similarities in temperature structure and bulk composition. However, there are some subtle differences that result in vertically more extended clouds and higher amount of gas above the cloud-bases on Saturn \citep{west_clouds_2009,atreya_comparison_1999}. Saturn and Neptune observations also show evidence of \ce{CH4} and \ce{H2S} clouds in addition to \ce{NH3}, \ce{NH4SH}, and \ce{H2O} clouds, due to their colder environments \citep{irwin_detection_2018,irwin_probable_2019}.

Although not a planet anymore, Pluto, as seen by the New Horizon spacecraft, is covered by layers of hazes \citep{gladstone_atmosphere_2016,west_planetary_2017}, likely due to condensation of hydrocarbons and nitriles such as \ce{C2H2}, \ce{C2H4}, \ce{C2H6}, and HCN \citep{gao_constraints_2017}. We also have to mention Titan, since it has a unique and substantial atmosphere, with hydrocarbon and nitrile cloud types. This includes HCN, \ce{HC3N}, \ce{CH4}, \ce{C2H2}, and \ce{C2H6} coated particles \citep{lavvas_condensation_2011}.

Clouds on gaseous planets beyond the solar system have also been observed. \citet{sing_continuum_2016} studied the transmission spectra of 10 Hot Jupiters and concluded that there is very likely a continuum of exoplanetary atmospheres; covered by haze and clouds with a variety of compositions. For instance, a nearly flat transmission spectrum of GJ\;1214b (a sub-Neptune planet with an equilibrium temperature of 580\;K) is consistent with a planet which is fully covered by clouds; most likely composed of ZnS and KCl \citep{kreidberg_clouds_2014}. Hotter Class-II gaseous exoplanets\footnote{See the first paper in this series for the classification scheme \citep{molaverdikhani_cold_2019}.} with a temperature range of $\sim$1000\;K to $\sim$1650\;K could have \ce{Na2S}, MnS, Fe, Al$-$oxides, or silicate clouds such as \ce{Mg2SiO3} and \ce{Mg2SiO4} clouds. Even Ultra-Hot Jupiters (T$\gtrsim$2500\;K) are expected to have cloud formation on their nightside \citep{helling_sparkling_2019,keating_universal_2018}, which could be dragged toward the dusk terminator of the planet and potentially contribute in the transmission spectra and transit depth of these planets (e.g. Helling in perp.). For a recent review on exoplanet cloud formation see \citet{helling_exoplanet_2019}.

%\note{Maybe show TPs of solar system and some well known exoplanets like HD189733b, HD209458b, GJ1124, GJ436 and saturation curves; from $https://www.researchgate.net/figure/The-vertical-temperature-profile-in-the-atmospheres-of-the-planets-is-shown-by-a_fig1_243492714$ and $https://www.researchgate.net/figure/Temperatures-versus-pressure-for-Venus-Earth-Mars-Jupiter-Saturn-Titan-Uranus_fig2_227050297$}

All these studies suggest the importance of condensation and particle formation in planetary atmosphere models. However, including even a simplified microphysiscs model, such as atmospheric models by \citet{rossow_cloud_1978}, in a chemical kinetic model demands high computational resources mostly because of the shorter timescales that condensation and evaporation operate on relative to the vapor transport or gas-phase chemistry timescales \citep{moses_chemical_2014}. Hence a simpler approach is usually taken. 

\citet{dobrijevic_key_2010} assumed condensation of species when their abundances reach their saturation level. After hitting this condition, they did not solve the continuity equation of that species at the given altitude; resulting in numerical stability and high computational efficiency of the scheme. A consequence of this approach is that the abundance of the condensed species follows its saturation curve perfectly. But condensation processes could occur at different saturation levels. Capillary condensation is an example, which through this mechanism the vapor condensation occurs below the saturation vapor pressure \citep{stokes_fundamentals_1997}.

\citet{moses_photochemistry_2000,moses_photochemistry_2000-1} treated the condensed phase of a species as a separate species that is produced by condensation and lost by evaporation through the following reactions:

\begin{equation}
\label{eq:cond}
\ce{species(g) + dust -> species(s) }
\end{equation}

\begin{equation}
\label{eq:V}
\ce{species(s) + V -> species(g) }
\end{equation}

where ``dust'' represents the condensation nuclei (CN) and must be provided, ``V'' represents a dummy molecule and its number density assumed to be 1 cm$^{-3}$ at all atmospheric levels, and the subscripts (g) and (s) refer to the gas and condensed phases, respectively. This scheme assumes the pre-existence of CN in the atmosphere; allowing the condensation of materials on these sites. They estimated the rate of these reactions by assuming steady-state diffusion-limited condensation and followed \citet{seinfeld_atmospheric_1986} rate estimations.

We follow a similar approach to account for the condensation, but with slightly different rate estimations adapted from \citet{seinfeld_atmospheric_2012} as follow:

\begin{equation}
\label{eq:rate}
J_c=4\pi R_p D_g (c_{\infty}-c_s)
\end{equation}

where $J_c$ is the total flow of the species toward the particle (moles sec$^{-1}$), $R_p$ is the particle's radius, $D_g$ is the diffusivity of the species in the atmosphere, $c_{\infty}$ is the concentration of the species far from the particle, and $c_s$ is its vapor-phase concentration at the particle surface. It is evident that when $c_{\infty}>c_s$ then the species flows toward the particle and if $c_{\infty}<c_s$ it moves away from it. $c_s$ can be estimated from the vapour pressure of species at any given temperature (vapour pressure data of more than 1700 substances are provided in \citet{haynes_crc_2016} and being updated regularly). To consider the condensation in \texttt{ChemKM}, the user is only required to provide a list of desired species to condense. The code then internally handles the addition of ``dummy'' reactions and condensed phase of species.

Gravitational acceleration leads to the settlement of condensed particles, i.e. rainout, which tends to remove the condensed materials from the atmosphere. Hence, not including the rainout could cause some discrepancies between the observations and models. Two well-known cases are the detection of \ce{H2S} gas on Jupiter \citep{niemann_composition_1998} and the presence of gaseous alkalies in cool brown dwarf and exoplanet atmospheres \citep{marley_clouds_2002,morley_neglected_2012,line_uniform_2017,zalesky_uniform_2019}. On Jupiter, Fe could deplete all sulphur content from the upper atmosphere through condensation of FeS; leaving no sulphur to form \ce{H2S}. Adding a Fe-rainout mechanism, however, could deplete all Fe from those regions, hence allowing \ce{H2S} to form. Similarly, on brown dwarfs and exoplanets, retaining the silicates at high altitudes can eventually lead to the formation of alkali feldspars; resulting in the depletion of alkalies from the upper atmosphere. Including silicate-rainout processes could prevent such an outcome by removing the silicates from those regions \citep[e.g.,][]{molliere_model_2015}. As an option in \texttt{ChemKM}, this effect can be included by removing the condensates as soon as they condense. A more sophisticated rainout scheme is under development to take into account a proper estimation of the drag force upon the particles and approximation of the settling velocity.

%%%%%%%%%%%%%%%%%%%%%%%%%%%%%%%%%%%%%%%%%%%%%%%%%%%%%%%%%%%
\subsection{Ablation and escape}\label{subsec:bc}

Ablation of micro-meteoroids during atmospheric entry is thought to be a potential source of volatile gases (such as \ce{H2O}, CO, \ce{CO2}, and \ce{SO2}) that cannot be maintained at the upper atmosphere of planets otherwise. These species could change the chemistry and dynamics of the upper layers. For instance, photolysis of the ablated water produces hydroxyl radicals (OH) \citep{moses_dust_2017}, which can react with many other species and cause new reaction pathways and products. Although the details of such processes (even on Earth) is not fully understood \citep[e.g.,][]{hawkes_what_2008}, usually an ablation rate for each species can be estimated based on the observations of the upper atmosphere of planets through a retrieval procedure \citep{moses_meteoroid_1992, horst_origin_2008,moses_dust_2017}.

Atmospheric escape is another process in action at the outermost layers of planetary atmospheres. Several atmospheric escape mechanisms have been identified so far; most of which are non-thermal processes, such as pickup by stellar-wind, bulk escape through magnetic flux ropes, or sputtering \citep[e.g.,][]{hunten_thermal_1982}. While these processes could be the dominant escape mechanisms on colder planets \citep[e.g.,][]{brain_spatial_2015,jakosky_initial_2015}, hydrodynamic escape (as one of the thermal processes) is believed to be, at least, as important as non-thermal mechanisms for hot exoplanets \citep[e.g.,][]{salz_simulating_2016}. Including all these processes into a chemical kinetic model would be a major task and computationally expensive. Thus, a simplified approach \citep[e.g.,][]{moses_seasonal_2018,dobrijevic_coupling_2014,hebrard_photochemistry_2013} could be useful. In this approximation, an escape rate is provided for each species at the top of the atmosphere (TOA) and species are removed from the uppermost layer by this rate.

%At the bottom boundary of terrestrial planets, outgassing and deposition can be in play and modify the chemical composition and temperature structure of planetary atmospheres. Volcanic eruptions can outgas \ce{SO2} and enhance formation of sulphur aerosols at high altitudes, which could remain in the stratosphere for months or even years \citep[e.g.,][]{hamill_analysis_1982,rampino_sulphur-rich_1984}. These aerosols can cool down the surface of Earth, in addition to the chemical changes that they cause at those atmospheric levels. Models of early Earth \citep[e.g.,][]{pavlov_mass-independent_2002,zahnle_impacts_2006} and early Mars \citep[e.g.,][]{plescia_assessment_1993} also suggest that significant volatile release and \ce{SO2} outgassing have changed the climates of both planets.

%Deposition of carbon and nitrogen compounds are well-known steps of the global carbon and nitrogen cycles on Earth \citep[e.g.,][]{falkowski_global_2000}. Measurements of Mars surface composition have also provided evidence of carbonate deposits at different locations \citep[e.g.,][]{hu_tracing_2015,niles_geochemistry_2013}. Similar to outgassing, deposition can also be accounted for through an estimation of its rate at the planetary surface, for each species.

Although gaseous planets have no solid surface, a virtual boundary can be defined as well. \revisiontwo{For the adequately hot planets, the bottom layer can be selected at a high pressure, where the atmosphere remains at thermochemical equilibirum. However, for cold gaseous planets,} their upper troposphere, where a cold-trap is formed, is usually selected to set a boundary layer condition for the chemical models \revisiontwo{as the bottom layer in the kinetic models is usually not at the thermochemical equilibrium and the local atmosphere is likely prone to the disequilibrium processes}. This setting provides more flexibility at the bottom layer of chemical models. For instance, it can be used to estimate the vertical \ce{CH4} vapor flux that makes it past the tropopause cold trap of Uranus and Neptune \citep[e.g.,][]{moses_seasonal_2018}.

\revision{These processes can be included by setting the flux rate of species at the top of the atmosphere. Similarly, outgassing and deposition can be included by setting the flux rate of species at the bottom boundary layer in \texttt{ChemKM}.}

%%%%%%%%%%%%%%%%%%%%%%%%%%%%%%%%%%%%%%%%%%%%%%%%%%%%%%%%%%%
\subsection{Cosmic rays and scattered Ly-alpha by LIPM}\label{subsec:gcr}

Galactic and solar cosmic rays (GCRs) are the main source of ionization at altitudes below 55–60 km on Earth \citep[][]{velinov_ionization_1968,bazilevskaya_observations_2000}; shaping Earth's stratosphere and thermosphere. GCRs are also the main driving factor of Titan's photochemistry at pressures between 1 and 50\;mbar. Particularly GCR-cascade dissociates \ce{N2}, a crucial process to synthesis nitrogen organics such as HCN \citep[e.g.,][]{capone_galactic_1983}. GCRs have shown to be important on Neptune's \citep[e.g.,][]{selesnick_neptunes_1991,lellouch_vertical_1994,aplin_determining_2016}, Titan's \citep[e.g.,][]{molina-cuberos_ionization_1999,yung_photochemistry_1984}, and likely exoplanets' \citep[e.g.,][]{helling_ionization_2011,rimmer_ionization_2013,scheucher_new_2018} upper atmospheres, too. A list of GCR reactions along with their rate profiles can be provided to the current version of \texttt{ChemKM} for the inclusion of GCRs. A new GCR module is under development to self-consistently calculate the GCR rates.

Isotropic source of stellar background UV radiation and solar Lyman-$\alpha$ photons that are scattered from atomic hydrogen in the local interplanetary medium (LIPM) are important dissociation/ionization sources at weakly irradiated environments \citep[e.g.,][]{strobel_nitrogen_1991,moses_i._1991,bishop_voyager_1998}, such as polar regions and nightside of closely orbiting planets or the entire atmosphere of planets orbiting at large distances from their host stars. A wavelength dependent flux of these irradiation sources can be provided to \texttt{ChemKM} to be included in the model.

%\note{Maybe show a schematic cartoon about the physics in ChemKM; i.e. mixing, photolysis by iradiation, by Ly-alpha glow, GCR, condensation and rainout, outgassing, escape, etc. with ChemKM's icon.}

%%%%%%%%%%%%%%%%%%%%%%%%%%%%%%%%%%%%%%%%%%%%%%%%%%%%%%%%%%%
\subsection{Initialization}\label{subsec:ic}

Once all desired physics in the model are set up, one should provide the initial conditions for the temperature-pressure (TP) and abundance profiles. The current version of \texttt{ChemKM} only considers the TP structure statically; i.e. the TP structure does not change as abundances evolve with time. Abundances can be initialized with two options: 1) by providing the elemental abundances or 2) by providing the molar fraction of any desired species at t$_0$=0.

As a first step, it is possible to calculate the thermochemical equilibrium abundances of species through Gibbs free minimization. We use \revision{a slightly modified version of} petitCODE's easy\_chem module to perform this \citep{molliere_observing_2017}. Alternatively, the thermochemical equilibrium can be achieved through kinetic calculations. The first approach is usually more suitable, unless the temporal evolution of a system to its thermochemical equilibrium state is as of interest.

Finally, depending on what physical processes are considered in the model, other inputs are also required; for instance the vertical mixing coefficient profile for the eddy diffusion or the influx rates.

%%%%%%%%%%%%%%%%%%%%%%%%%%%%%%%%%%%%%%%%%%%%%%%%%%%%%%%%%%%
\begin{figure}[t]
\includegraphics[width=\columnwidth]{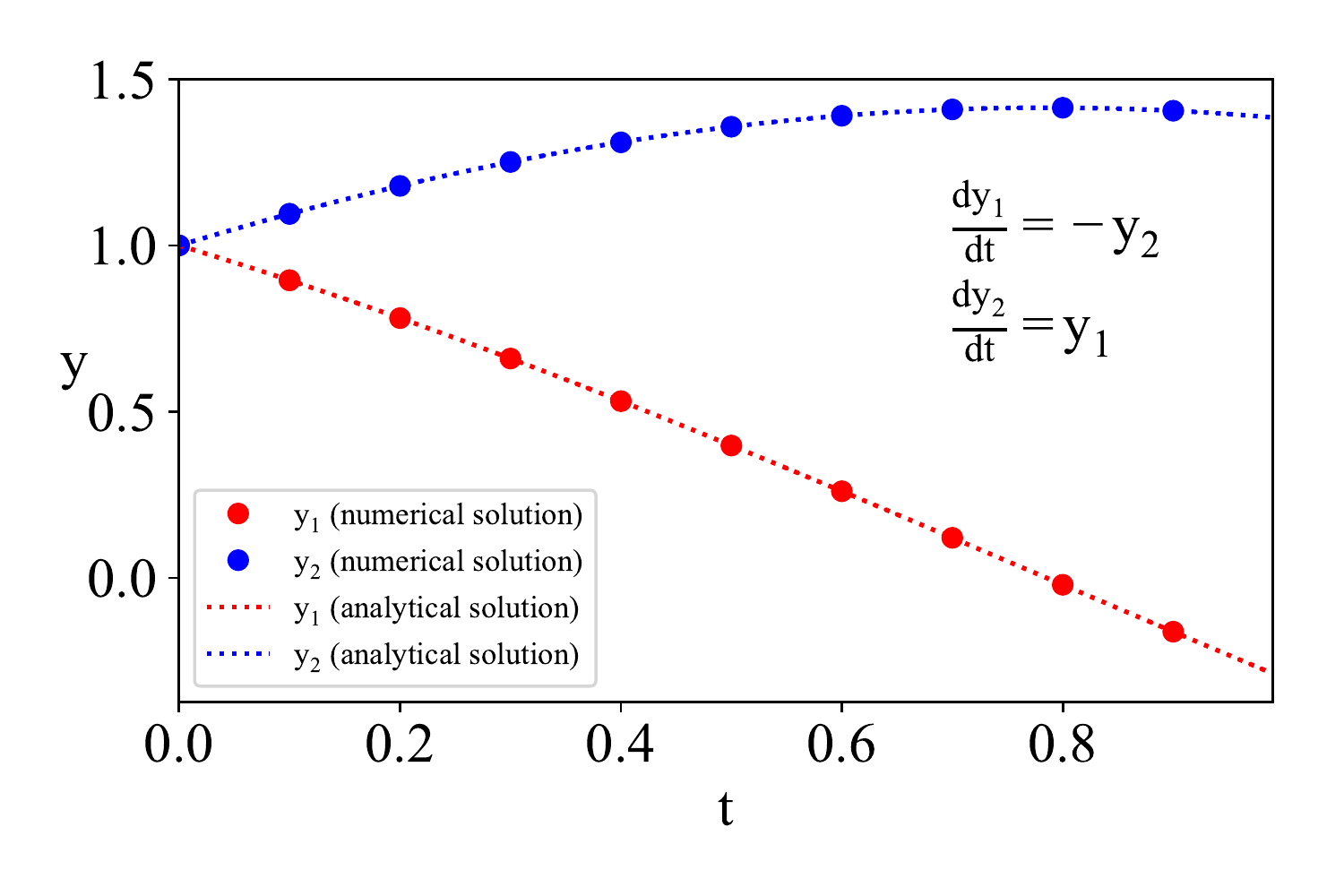}
\caption{Numerical integrator's performance for a simple case. The results are compared with the analytical solution of the system. They agree within the numerical errors. \label{fig:analytical}}
\end{figure}

%%%%%%%%%%%%%%%%%%%%%%%%%%%%%%%%%%%%%%%%%%%%%%%%%%%%%%%%%%%
%%%%%%%%%%%%%%%%%%%%%%%%%%%%%%%%%%%%%%%%%%%%%%%%%%%%%%%%%%%
%%%%%%%%%%%%%%%%%%%%%%%%%%%%%%%%%%%%%%%%%%%%%%%%%%%%%%%%%%%
\section{Model verification} \label{sec:verification}

%%%%%%%%%%%%%%%%%%%%%%%%%%%%%%%%%%%%%%%%%%%%%%%%%%%%%%%%%%%
\subsection{Numerical integrator} \label{subsec:integrator}

We start the verification of the code by showing how the DLSODE numerical integrator performs.  \hyperref[fig:analytical]{Figure}\;\ref{fig:analytical} shows the numerical and analytical solutions of the following system of ODEs:

\begin{equation}
\label{eq:analytical}
\begin{split}
\frac{dy_1}{dt}=-y_2 \\ \frac{dy_2}{dt}=y_1
\end{split}
\end{equation}

where $y_i$ is the abundance of species $i$ at any given time, $t$. Abundance variation of each species depends on one another, hence the system is coupled. The analytical solution of this simple system is as follow, assuming $y_1(0)=y_2(0)=1.0$:

\begin{equation}
\label{eq:analytical}
\begin{split}
y_1=Cos(t)-Sin(t) \\ y_2=Cos(t)+Sin(t)
\end{split}
\end{equation}

The results of analytical and numerical solutions are in agreement, hence the numerical performance is assured within the numerical errors.

%%%%%%%%%%%%%%%%%%%%%%%%%%%%%%%%%%%%%%%%%%%%%%%%%%%%%%%%%%%
\subsection{Thermochemical equilibrium} \label{subsec:Validation}

%%%%%%%%%%%%%%%%%%%%% reversable_reac_0D
\begin{figure}
\includegraphics[height=\dimexpr \textheight - 4\baselineskip\relax]{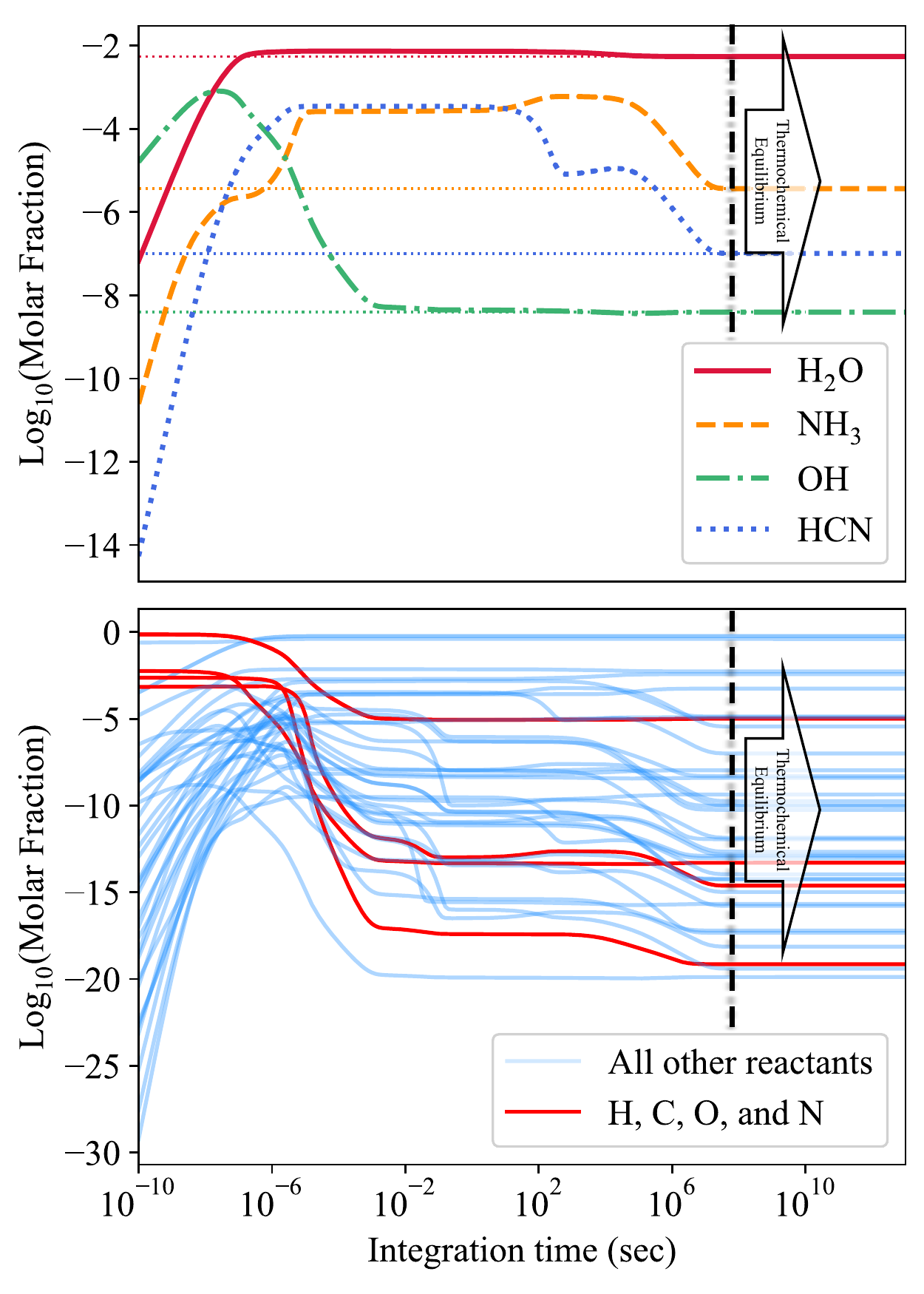}
\caption{Consistency of the thermodynamic equilibrium composition (dotted straight lines) and \texttt{ChemKM}'s kinetic steady state solutions in a reversible system of reactants in a 0D model setup with T=1560\;K and P=10\;bar. The result of the thermochemically reversed model is shown as a function of integration time for different species. \textbf{Top)} Evolution of \ce{H2O} (solid red), \ce{NH3} (dashed orange), \ce{OH} (dashed-dotted green), and \ce{HCN} (dotted blue), until they reach their thermochemical equilibrium abundances. \textbf{Bottom)} Evolution of \ce{H2},	\ce{He},	\ce{H},	\ce{CO},	\ce{H2O},	\ce{CH4},	\ce{N2},	\ce{NH3},	\ce{CO2},	\ce{HCN},	\ce{C2H2},	\ce{O2},	O($^3$P),	O($^1$D),	\ce{C},	N($^4$S),	N($^2$D),	\ce{OH},	\ce{CH3},	$^3$\ce{CH2},	$^1$\ce{CH2},	\ce{CH},	\ce{NH2},	\ce{NH},	\ce{H2CO},	\ce{HCO},	\ce{CN},	\ce{C2H},	\ce{NO},	\ce{HNO},	\ce{NCO},	\ce{HNCO},	\ce{HOCN},	\ce{HCNO},	\ce{C2H3},	\ce{C2H4},	\ce{C2H5},	and \ce{C2H6} abundances. Atomic abundances (red lines) deplete as molecules (blue lines) are being produced under this condition. \label{fig:reversablereac0D}}
\end{figure}
%%%%%%%%%%%%%%%%%%%%%

Because of the relatively high temperatures at the photosphere of most of known giant exoplanets, simulations of their atmospheric chemistry usually require the consideration of reversible reactions in the model. One issue that have been addressed by previous studies as well (e.g. \citet{moses_disequilibrium_2011,venot_chemical_2012,drummond_effects_2016}) is the unavailability of rate coefficients for most reverse reactions. A common workaround is to calculate these rates as the ratio between the forward rate constant and the equilibrium constant. The equilibrium constant, therefore, can be calculated by using NASA thermodynamic polynomial coefficients \citep{mcbride_nasa_2002}. Inclusion or exclusion of reverse reactions in \texttt{ChemKM} is optional, but if it is chosen to be included then it follows the above described method to calculate the reverse rate coefficients.

Taking this approach ensures the consistency of kinetics and thermodynamics steady state solutions. \hyperref[fig:reversablereac0D]{Figure}\;\ref{fig:reversablereac0D} shows this consistency and compares the results of a simple 0D model at T=1560\;K and P=10\;bar for a kinetic chemical model with reversible reactions (curves) and thermodynamic equilibrium calculations (straight thin dotted lines). The kinetic model reaches the thermodynamic equilibrium after 10$^7$\;sec with molar fractions practically being identical. Initial atomic abundances (red lines) deplete as molecules (blue lines) are being produced. Obviously, the chemical evolution of reactants depends on the choice of chemical network. For this test, and all other verification tests in this section, we employ the \citet{venot_chemical_2012} kinetic network as a verified and benchmarked chemical network.

\begin{figure*}
\includegraphics[width=\textwidth]{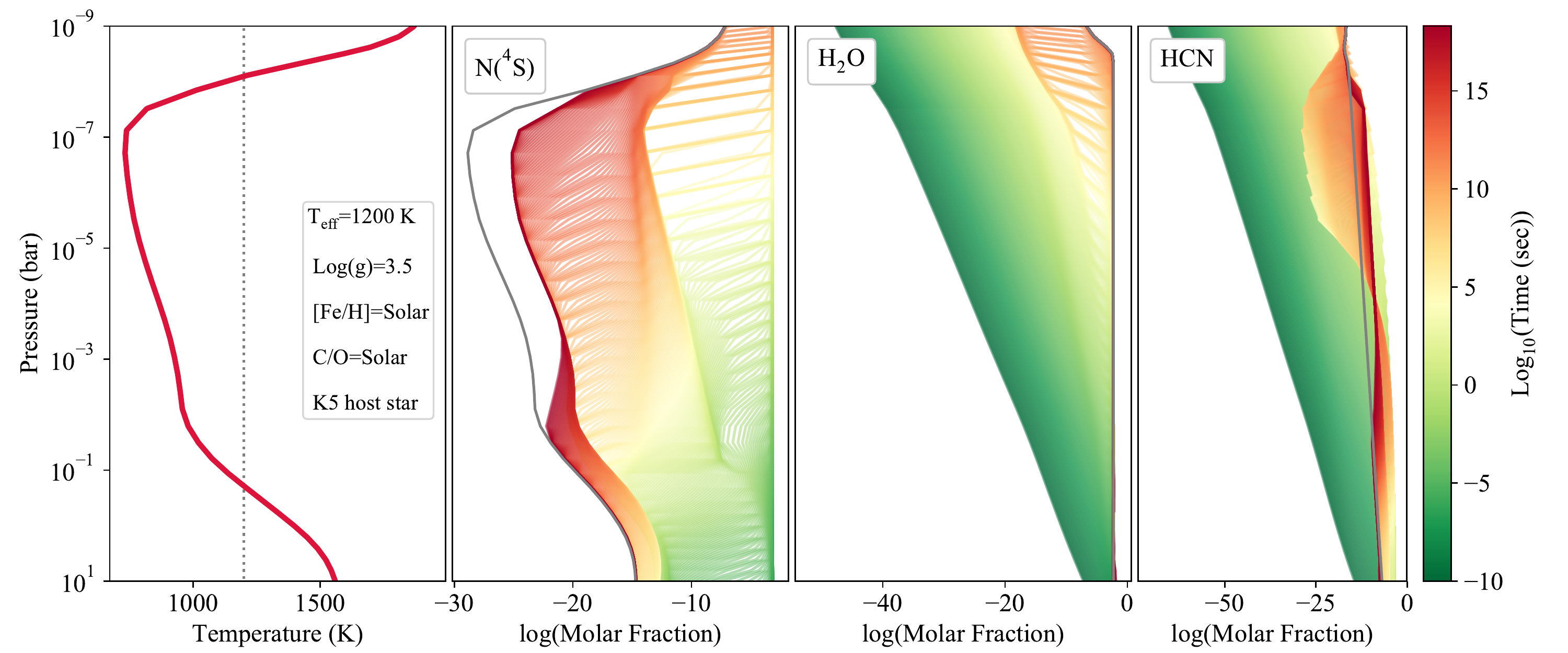}
\caption{Similar to \hyperref[fig:reversablereac0D]{Figure}\;\ref{fig:reversablereac0D} but with a 1D model; without the inclusion of vertical mixing or photolyses. \textbf{left)} Temperature profile of an irradiated exoplanet with T\textsubscript{eff}$=$1200\;K, log(g)$=$3.5, with solar metallicity and C/O ratio, orbiting around a K5 star. \textbf{Rest of panels)} from left to right: The temporal evolution of N($^4$S), \ce{H2O}, and HCN abundances, color-coded by time, where green to red shows the progress in time from 10$^{-10}$\;sec to more than the age of universe, 1.6$\times$10$^{18}$\;sec. The thermochemically reversed model (colored lines) is in agreement with the thermochemical equilibrium calculations (gray lines) at adequately high temperatures ($\geq$1000\;K). However, abundances at pressures between 10$^{-2}$ and 10$^{-8}$\;bar ($<$1000\;K) have not fully reached their thermochemical equilibrium yet and demand longer integration times.
\label{fig:reversreac1D}}
\end{figure*}

\hyperref[fig:reversreac1D]{Figure}\;\ref{fig:reversreac1D} provides a similar comparison for a 1D model. Neither our 0D nor the 1D model includes mixing or photolyses for these two tests. The temporal evolution of abundances are color-coded by time where green to red shows the progress in time from 10$^{-10}$\;sec to 1.6$\times$10$^{18}$\;sec (beyond the age of universe). We show a chemically important atom, N($^4$S), one oxygen$-$bearing major opacity molecule, \ce{H2O}, and one expectedly abundant carbon$-$bearing molecule at high temperatures, HCN, as examples. It is evident that the abundances at kinetically steady state (dark red) are in agreement with their thermodynamic equilibrium values (gray lines) when the local temperature is adequately high (usually above 1000\;K). The N($^4$S) and HCN abundances at pressures between 10$^{-2}$ and 10$^{-8}$\;bar have not fully reached their thermochemical equilibrium and require longer integration times. This shows that the calculation of thermochemical equilibrium is better performed through Gibbs free energy minimization.

%%%%%%%%%%%%%%%%%%%%%
%%%%%%%%%%%%%%%%%%%%%

As already mentioned, the ``kinetically thermochemical equilibrium'' mode, i.e. solving the system of ODEs without diffusion or photolyses, can be used as a pre-processing step to initialize the model with thermochemical equilibrium abundances. However, \texttt{ChemKM} has the option to quickly calculate abundances at thermochemical equilibrium by using Gibbs free energy minimization. \hyperref[fig:rtols]{Figure}\;\ref{fig:rtols} shows an example of such a setup, where the model is initialized by using Gibbs free energy minimization and {\it all} abundances remain the same in the kinetic calculations due to the lack of additional physics such as diffusion or photolyses, hence verifying the validity of our reversible setup to reach or keep the same composition as in thermochemical equilibrium. See \hyperref[subsec:ic]{Section}\;\ref{subsec:ic} for a description of initialization options in ChemKM.

%While a thorough benchmark with other widely used kinetic chemical models is in progress [Drummond et al. in prep.], the following sections present some case studies using the \texttt{ChemKM} to study the effects of different physics at different atmospheric conditions.

%%%%%%%%%%%%%%%%%%%%%%%%%%%%%%%%%%%%%%%%%%%%%%%%%%%%%%%%%%%
\subsection{Molecular and eddy diffusion}  \label{subsec:eddy}

Transport-induced quenching is one of the main kinetic-related disequilibrium processes in planetary atmospheres. Both molecular and eddy diffusion could play pivotal roles to drive away the atmospheric composition from their thermochemical equilibrium. If the chemical kinetic timescales are larger than the transport timescales, the mole fraction of a parcel of gas can become quenched in the atmosphere. This usually happens when the temperature and pressure are low enough that the kinetic reactions cannot rapidly occur in both directions; allowing the diffusion processes to drive the system away from its thermochemical equilibrium state.

\begin{figure*}
\includegraphics[width=\textwidth]{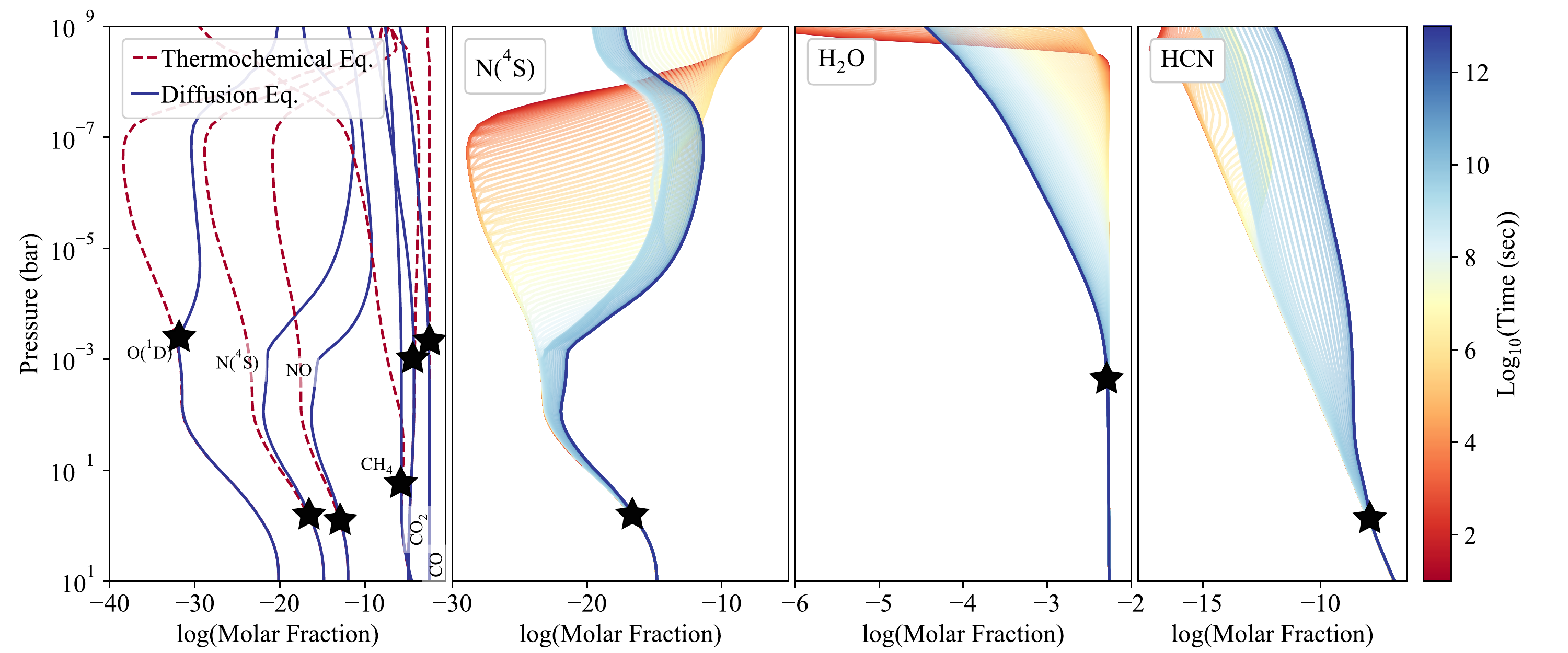}
\caption{Time evolution of our 1D model presented in \hyperref[fig:reversreac1D]{Figure}\;\ref{fig:reversreac1D} from its thermochemical equilibrium state to diffusion equilibrium, by introducing molecular diffusion and a vertically constant eddy diffusion (K\textsubscript{zz}=$10^{4}$\;cm$^2$s$^{-1}$). \textbf{Left panel)} Thermochemical equilibrium (dashed red lines) and diffusion equilibrium of O($^1$D), N($^4$S),\ce{NO, CH4, CO2, CO} (solid blue lines). \textbf{Rest of panels)} Temporal evolution of N($^4$S), \ce{H2O}, and \ce{HCN} abundances in detail. They are color-coded by time, where blue to red shows the progress in time from 10$^{1}$\;sec to 10$^{13}$\;sec. \revision{The quenching points are shown by the stars.} \label{fig:reversreac1Ddiff}}
\end{figure*}

\begin{figure}
\includegraphics[height=\dimexpr \textheight - 4\baselineskip\relax]{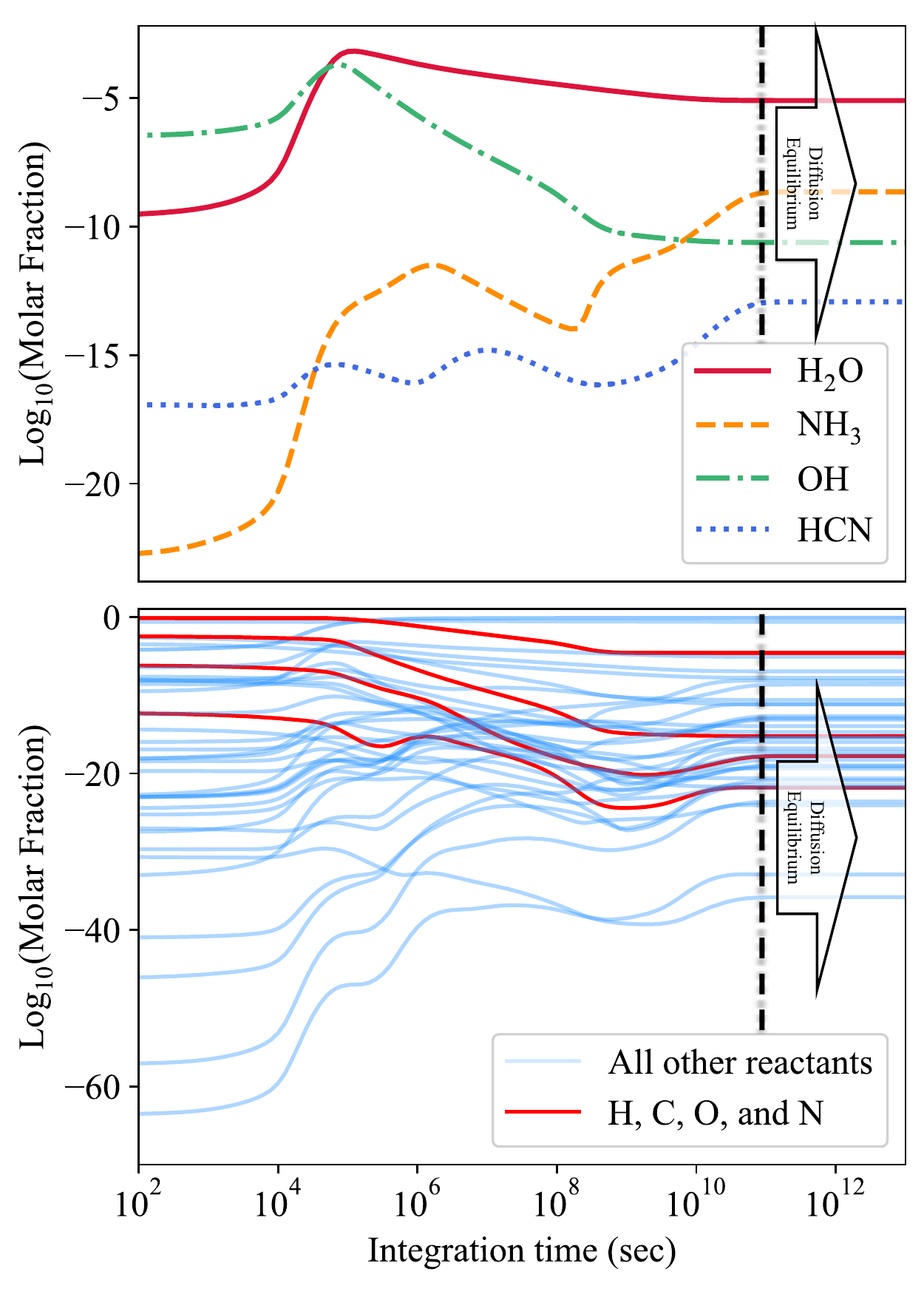}
\caption{A cross-section of \hyperref[fig:reversreac1Ddiff]{Figure}\;\ref{fig:reversreac1Ddiff} at 10$^{-10}$\;bar to illustrate the "diffusion-equilibrium" at the top of the atmosphere after 10$^{11}$\;sec. \label{fig:reversreac0Ddiff}}
\end{figure}

\hyperref[fig:reversreac1Ddiff]{Figure}\;\ref{fig:reversreac1Ddiff} gives an example for such a system with a 1D atmosphere evolving from its thermochemical equilibrium state. We use the same setup as our 1D example presented in the previous section (\hyperref[fig:reversreac1D]{Figure}\;\ref{fig:reversreac1D}), but initialize the composition with their thermochemical equilibrium values. We include a vertically constant eddy diffusion (K\textsubscript{zz}=$10^{4}$\;cm$^2$s$^{-1}$) in the model. We choose this relatively low K\textsubscript{zz} to demonstrate its significance even in the case of a hot atmosphere (e.g. in this case T\textsubscript{eff}=1200\;K). No photolyses is included in this setup to only present the effect of diffusion. The abundances reach to a new steady state known as the ``diffusion equilibrium'' \cite[e.g.][]{lettau_diffusion_1951}. A cross-section of \hyperref[fig:reversreac1Ddiff]{Figure}\;\ref{fig:reversreac1Ddiff} at 10$^{-10}$\;bar is shown in \hyperref[fig:reversreac0Ddiff]{Figure}\;\ref{fig:reversreac0Ddiff} to graphically illustrate the evolution of abundances until they reach the diffusion equilibrium at TOA. Continuing the integration until 10$^{20}$\;sec did not change this new steady state within numerical errors; hence we conclude that this state is a real diffusion equilibrium.

%\footnote{We also show that including other disequilibrium processes in a model results in a new steady state as well and the system remains at an ``equilibirum'' afterward. Hence, explicitly naming the achieved equilibrium state (such as ``thermochemical equilibrium'', ``diffusion equilibrium'', and ``photochemical equilibrium'') seem to be more appropriate than out-of-equilibrium or non-equilibrium.}

%%%%%%%%%%%%%%%%%%%%%
%%%%%%%%%%%%%%%%%%%%%

%%%%%%%%%%%%%%%%%%%%%
%%%%%%%%%%%%%%%%%%%%%

The quenching levels of of O($^1$D), N($^4$S),\ce{NO}, \ce{CH4}, \ce{CO2}, \ce{CO}, see \hyperref[fig:reversreac1Ddiff]{Figure}\;\ref{fig:reversreac1Ddiff}, are in agreement with the general estimations of quenching levels in  \citet{venot_better_2018}. The quenching level of each constituent, however, depends on its chemical kinetics timescales, ranging from 1 to 10$^{-3}$\;bar in this example.

A commonly used definition of the ``quenching point'' (a.k.a. quenching pressure or quenching level) is that the composition of all species remain constant above that level, that is the abundances remain the same as their abundances of their quenching point \cite[e.g.][]{moses_chemical_2014}. This assumption has been also used to mimic a parameterization of disequilibrium processes, usually in the retrieval models as a trade-off to gain more speed \citep[e.g.][]{madhusudhan_high_2011}. However, a constant abundance profile is rarely the case in realistic planetary atmospheres and atmospheric constituent abundances could show significant deviations from their quenched abundances. Prominent processes at mbar level, such as molecular diffusion or photolysis, could drive the abundance profiles away from an idealized constant profile. A non-isothermal TP structure, for example a hot thermosphere with local heating at TOA (as it is the case in the current example, see the TP structure in \hyperref[fig:reversreac1D]{Figure}\;\ref{fig:reversreac1D} ), can result in a variation of chemical timescales at different altitudes and might trigger the system to move in and out of thermochemical equilibrium. Including the vertical diffusion causes mixing of these abundances at different pressures; potentially keeping them away from a constant profile. For instance, dissociation of \ce{H2O} at TOA due to a hot upper thermosphere depletes its abundances at higher pressures through vertical mixing, see \hyperref[fig:reversreac1Ddiff]{Figure}\;\ref{fig:reversreac1Ddiff} and \hyperref[fig:gCVexample]{Figure}\;\ref{fig:gCVexample}.

In addition, such sharp temperature gradient could also enhance the molecular diffusion (see \hyperref[eq:moldiff]{Equation}\;\ref{eq:moldiff}), resulting in more deviation from a constant abundance profile. The left panel of \hyperref[fig:reversreac1Ddiff]{Figure}\;\ref{fig:reversreac1Ddiff} represents such examples, where abundances do not remain constant above their quenching points \revision{(black stars)} both due to molecular diffusion and thermal$-$dissociation. More discussion will follow in \hyperref[subsec:Quenching]{Section}\;\ref{subsec:Quenching}. As a rule of thumb, a constant quenched abundance profile is likely to be valid when the chemical kinetics timescale is much shorter than the timescale of all other processes at all pressures above the quenching point.

%%%%%%%%%%%%%%%%%%%%%%%%%%%%%%%%%%%%%%%%%%%%%%%%%%%%%%%%%%%
\subsection{Photochemistry} \label{sec:Photochemistry}

Photochemistry is another prominent disequilibrium process in planetary atmospheres. We examine the same case as in the previous section, but include photochemistry instead of diffusion. We use an updated version of \citet{hebrard_neutral_2012} photolysis reactions (originally adapted from the MPI-Mainz UV/VIS Spectral Atlas \citet{keller-rudek_mpi-mainz_2013}) and a stellar flux similar to the average solar flux over an entire solar cycle from \citet{thuillier_solar_2004} as the radiation flux at TOA. We initialize abundances with their thermochemical equilibrium values, identical to the previous section initialization.

\begin{figure*}
\includegraphics[width=\textwidth]{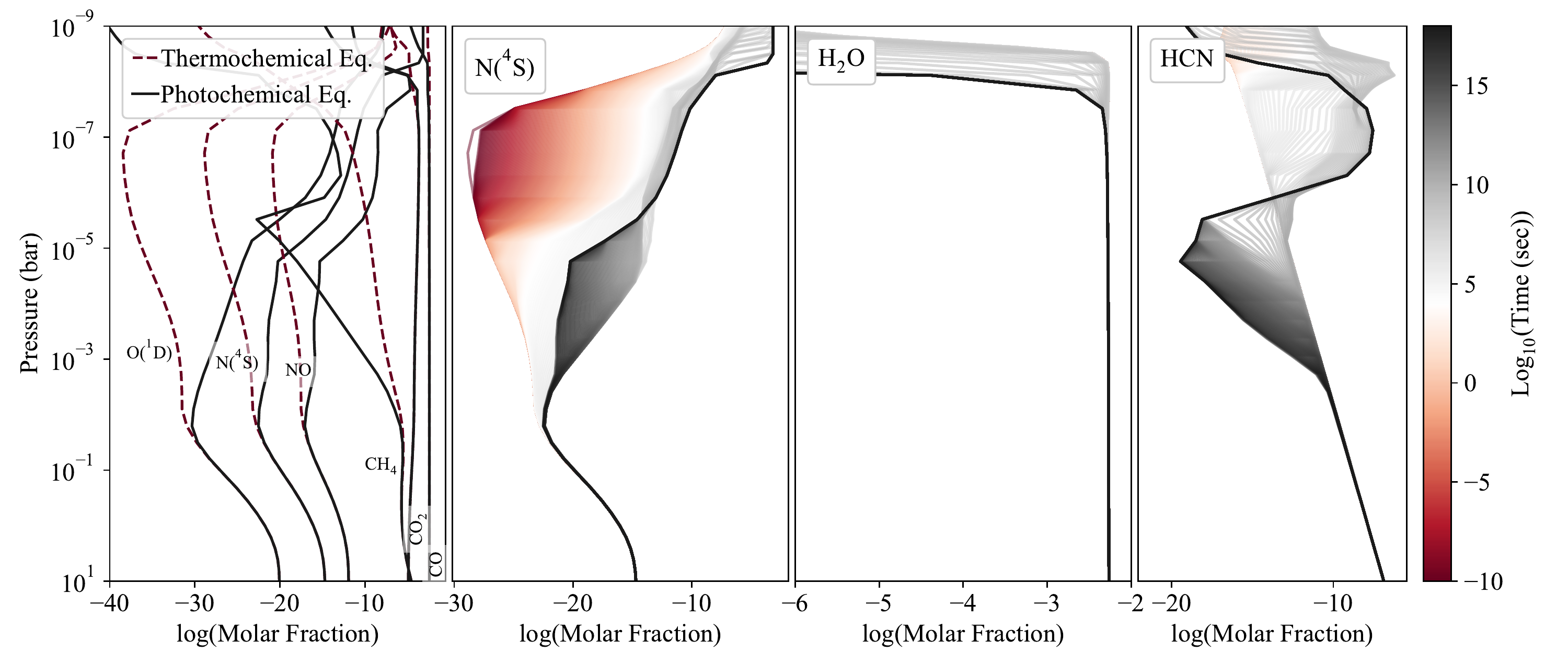}
\caption{Temporal evolution of the 1D model presented in \hyperref[fig:reversreac1D]{Figure}\;\ref{fig:reversreac1D} but evolving from its thermochemical equilibrium to photochemical equilibrium by introducing photolysis into the model. No diffusion is included in the model to purely present the effect of photolysis. The temporal evolution of abundances are  color-coded by time, where red to black shows the progress in time from 10$^{-10}$\;sec to beyond the age of universe, 10$^{18}$\;sec. Abundances around 10$^{-3}$\;bar have not fully reached their photochemical equilibrium yet and require longer integration times. \label{fig:reversreac1Dphoto}}
\end{figure*}

\hyperref[fig:reversreac1Dphoto]{Figure}\;\ref{fig:reversreac1Dphoto} shows the effect of photolysis on this model. The photochemical timescales are typically shorter than diffusion processes and hence the photolysis should be the dominant mechanism at TOA. Each species responds to the irradiation differently, depending on its shielding and UV cross-section. Some have excess production through photolysis reactions, e.g. N($^4$S) in this model, some face considerable destruction by this process, e.g. \ce{H2O}, and some could have a combination of those depending on the altitude, e.g. HCN.

After some time, the system reaches a new steady state, the ``photochemical equilibrium'', balancing between reversible reactions and photodissociation. The required integration time to achieve photochemical equilibrium at all vertical levels strongly depends on the atmospheric temperature structure; with hotter atmospheres tending to reach the photochemical equilibrium faster. This is evident in \hyperref[fig:reversreac1Dphoto]{Figure}\;\ref{fig:reversreac1Dphoto}, where reaching the photochemical equilibrium at around 10$^{-3}$\;bar requires integration times larger than the age of the universe.

\begin{figure}
\includegraphics[width=\textwidth]{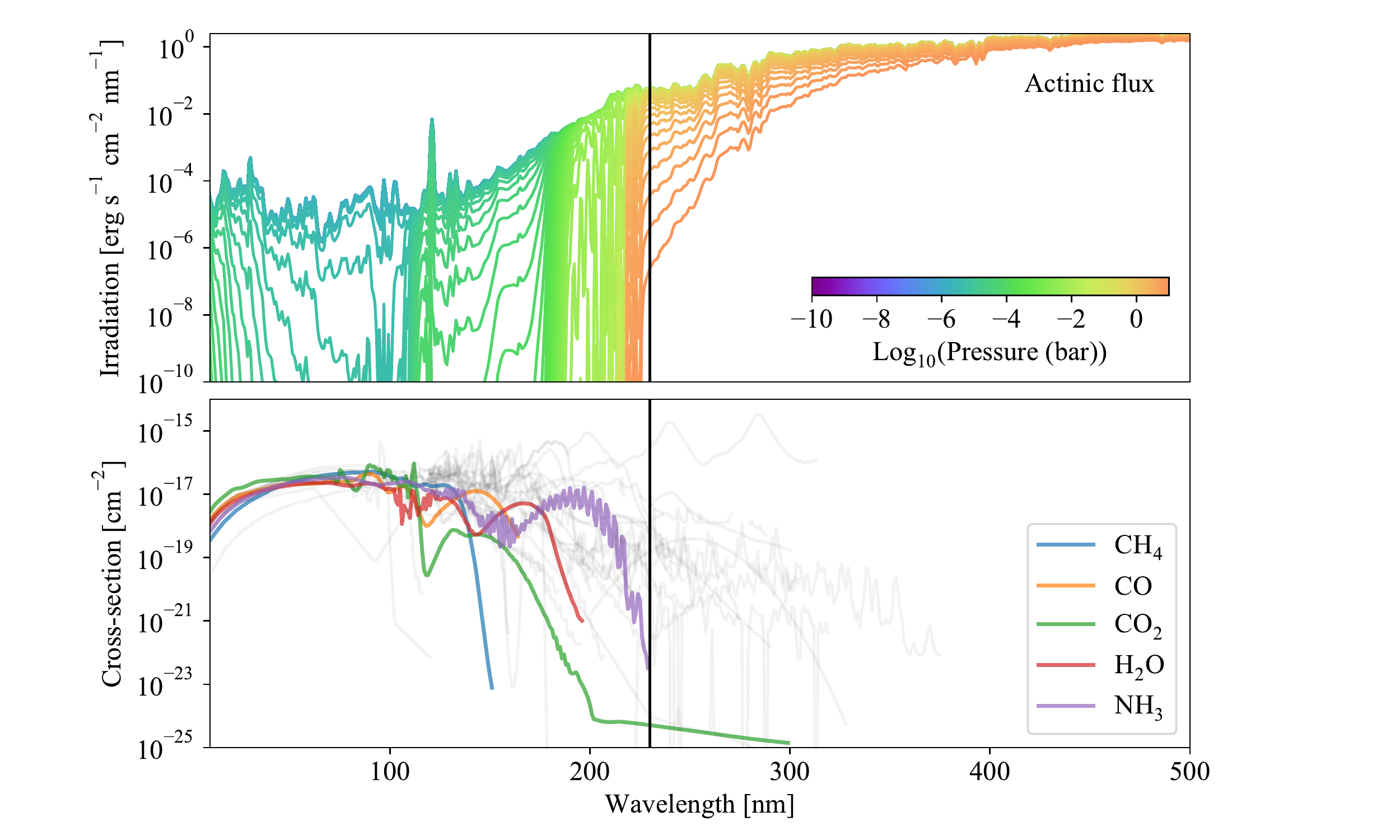}
\caption{ {\bf Top)} Local stellar intensity (actinic flux) in logarithmic scale. Lower pressures are color-coded by cooler colors. {\bf Bottom)} UV absorption cross-section of species in \texttt{ChemKM} (gray). Some of the species with usually major opacity contributions are shown in color.  \label{fig:fluxaltitude}}
\end{figure}

Absorption cross-sections, the quantum yields, and the actinic flux\footnote{``The quantity of light available to molecules at a particular point in the atmosphere and which, on absorption, drives photochemical processes in the atmosphere.'' See \citet{calvert_glossary_1990} for more details.} at a given altitude determine the photodissociation rate of a photo-reactant. \hyperref[fig:fluxaltitude]{Figure}\;\ref{fig:fluxaltitude} shows the irradiation at TOA and the actinic flux at different pressures. The stellar flux quickly vanishes at wavelengths shorter than 220\;nm due to strong UV absorption cross-section of photo-reactants at those wavelengths. The rest of the stellar flux is also affected by Rayleigh scattering in this model.

%%%%%%%%%%%%%%%%%%%%%
%%%%%%%%%%%%%%%%%%%%%

%%%%%%%%%%%%%%%%%%%%%
%%%%%%%%%%%%%%%%%%%%%

%%%%%%%%%%%%%%%%%%%%%%%%%%%%%%%%%%%%%%%%%%%%%%%%%%%%%%%%%%%
\subsection{Condensation} \label{sec:Condensation}

\begin{figure}
\includegraphics[width=\columnwidth]{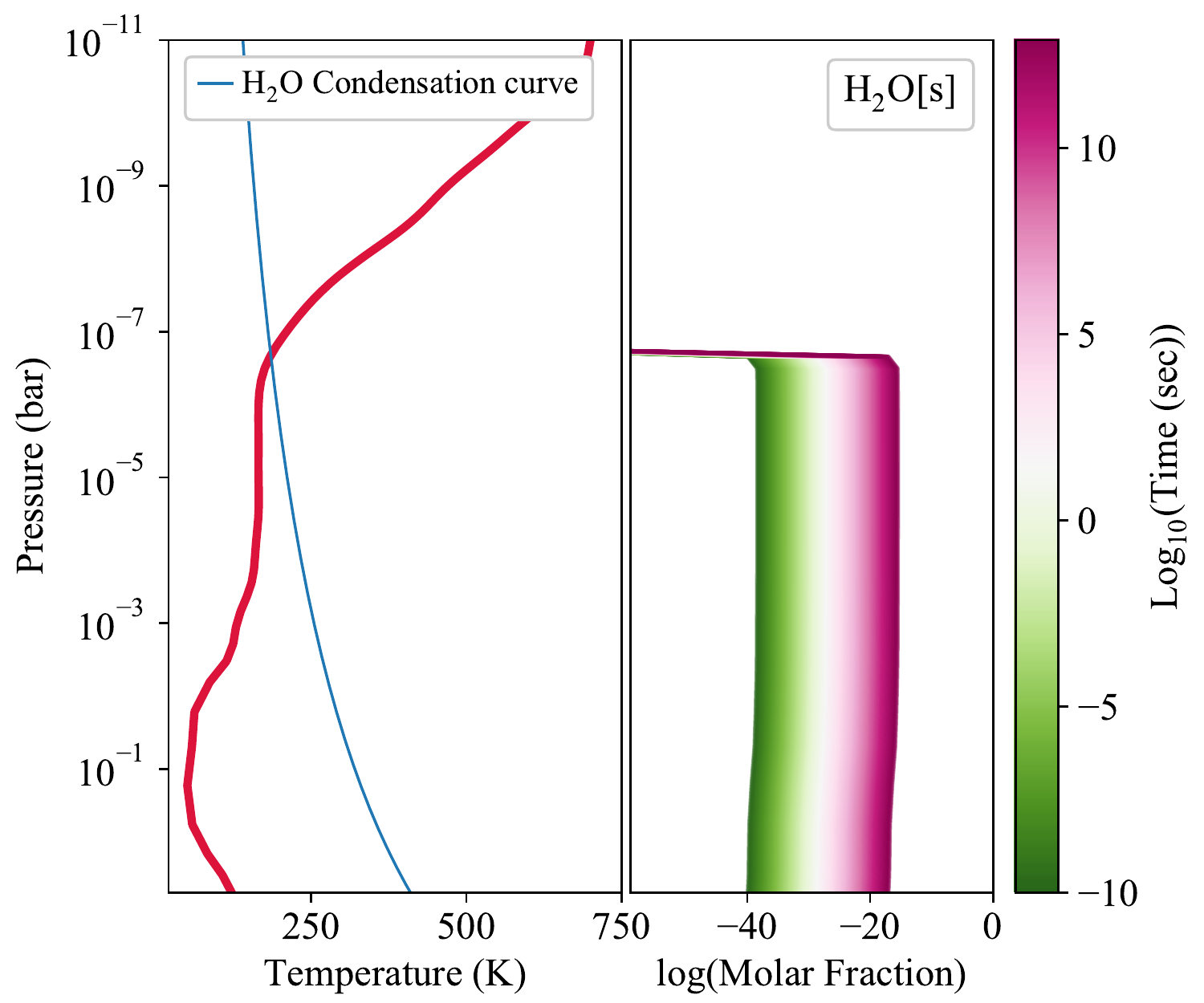}
\caption{Water condensation in a nearly water-saturated Neptune-like planet, i.e. the temperature profile is similar to that of Neptune, but the initial composition is set to 80.8$\%$ \ce{H2O}, 19$\%$ \ce{He}, and the rest is \ce{CH4} and \ce{CO} with molar fraction of $1.2\times10^{-3}$ and $8\times10^{-8}$ respectively. Water condensation occurs (right panel) below the intersection of the TP profile (left panel red line) and \ce{H2O}'s condensation curve (left panel blue line).  \label{fig:watercondsaturated}}
\end{figure}

To demonstrate \texttt{ChemKM}'s ability to capture condensation, we model an atmosphere with a temperature structure similar to that of Neptune. This ``cold'' case creates a suitable environment to examine this process, although \texttt{ChemKM} is capable of including the condensation for hot planets as well. No diffusion or photolysis is included in this model. We also employ the \citet{venot_photochimie_2012} kinetic network with no reverse reaction. First, we consider a nearly water-saturated atmosphere. This nearly water-saturated atmosphere would allow the condensation to begin at pressures below the intersection of Neptune's TP structure and \ce{H2O}'s condensation curve. We initialize the model identical to the \citet{moses_seasonal_2018} Neptune model, except for \ce{He} and \ce{H2O} abundances, and assume an initial atmospheric composition of 80.8$\%$ \ce{H2O}, 19$\%$ \ce{He}, $1.2\times10^{-3}$ \ce{CH4} and $8\times10^{-8}$ \ce{CO}. The particle size is also assumed to be 0.15 $\mu$m and fixed, following the \citet{moses_seasonal_2018} assumption. \hyperref[fig:watercondsaturated]{Figure}\;\ref{fig:watercondsaturated} shows the production of \ce{H2O}[s] (water ice) under these circumstances, with the expected condensation rates discussed in \hyperref[subsec:condrainout]{Section}\;\ref{subsec:condrainout}.

%%%%%%%%%%%%%%%%%%%%%
%%%%%%%%%%%%%%%%%%%%%

\begin{figure*}
\includegraphics[width=\textwidth]{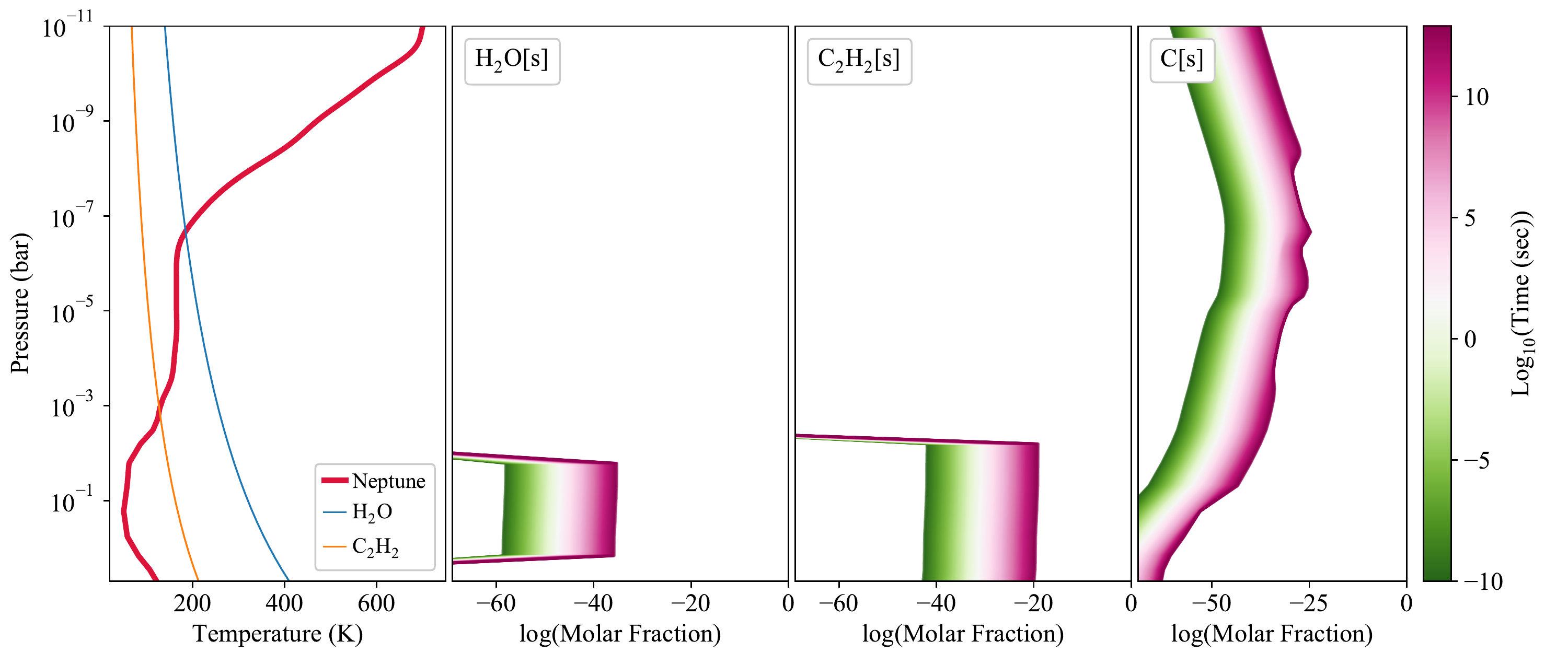}
\caption{Condensation and cold-trap on Neptune. {\bf Left)} Neptune's temperature structure (red line) and condensation curves of \ce{H2O} (blue line) and \ce{C2H2} (orange line). {\bf Rest of panels)} Temporal evolution of condensate production, and formation of a cold-trap roughly between 1 and 0.01\;bar. \label{fig:Neptunecond}}
\end{figure*}

In reality, water condensation in the atmosphere of Neptune occurs at higher pressures ($\approx$0.01-1\;bar) due to water undersaturation. However, the temperature increases again at the regions deeper than 1\;bar, causing the water to remain in the gas phase and making the $\approx$0.01-1\;bar region a tropopause ``cold-trap". This is shown for water condensation in \hyperref[fig:Neptunecond]{Figure}\;\ref{fig:Neptunecond}. The initial abundances for this model are 80.8$\%$ \ce{H2}, 19$\%$ \ce{He}, and the rest is \ce{CH4} and \ce{CO} with molar fraction of $1.2\times10^{-3}$ and $8\times10^{-8}$ respectively, identical to the initial conditions in \citet{moses_seasonal_2018}. We allow for the condensation of \ce{H2O}, \ce{NO}, \ce{N2}, \ce{O2}, \ce{C}, \ce{HCN}, \ce{CH4}, \ce{CO}, \ce{CO2}, \ce{C2H2}, \ce{C2H4}, and \ce{C2H6} for demonstration purposes. Condensation of \ce{C2H2} and C are shown in \hyperref[fig:Neptunecond]{Figure}\;\ref{fig:Neptunecond} as examples. In this setup, carbon condenses at all altitudes. However, most Neptune photochemical models do not allow carbon to condensate. This is to maintain a higher production rate of hydrocarbons, such as \ce{C2H2} and \ce{C2H6}.

%%%%%%%%%%%%%%%%%%%%%
%%%%%%%%%%%%%%%%%%%%%

%K. Willacy1, M. Allen2,3, and Y. Yung3 2016 also provides a table of Saturated Vapor Pressures

%%%%%%%%%%%%%%%%%%%%%%%%%%%%%%%%%%%%%%%%%%%%%%%%%%%%%%%%%%%
\subsection{Influxes and boundary conditions} \label{sec:influxes}

Ablation, atmospheric escape, and GCR can be included in \texttt{ChemKM} by setting the influx or production rate of reagents. To demonstrate these capabilities, we use the same Neptune model and initialization as the previous section. We include molecular and eddy diffusion as well as photolysis in this case.
 
\begin{figure*}
\includegraphics[width=\textwidth]{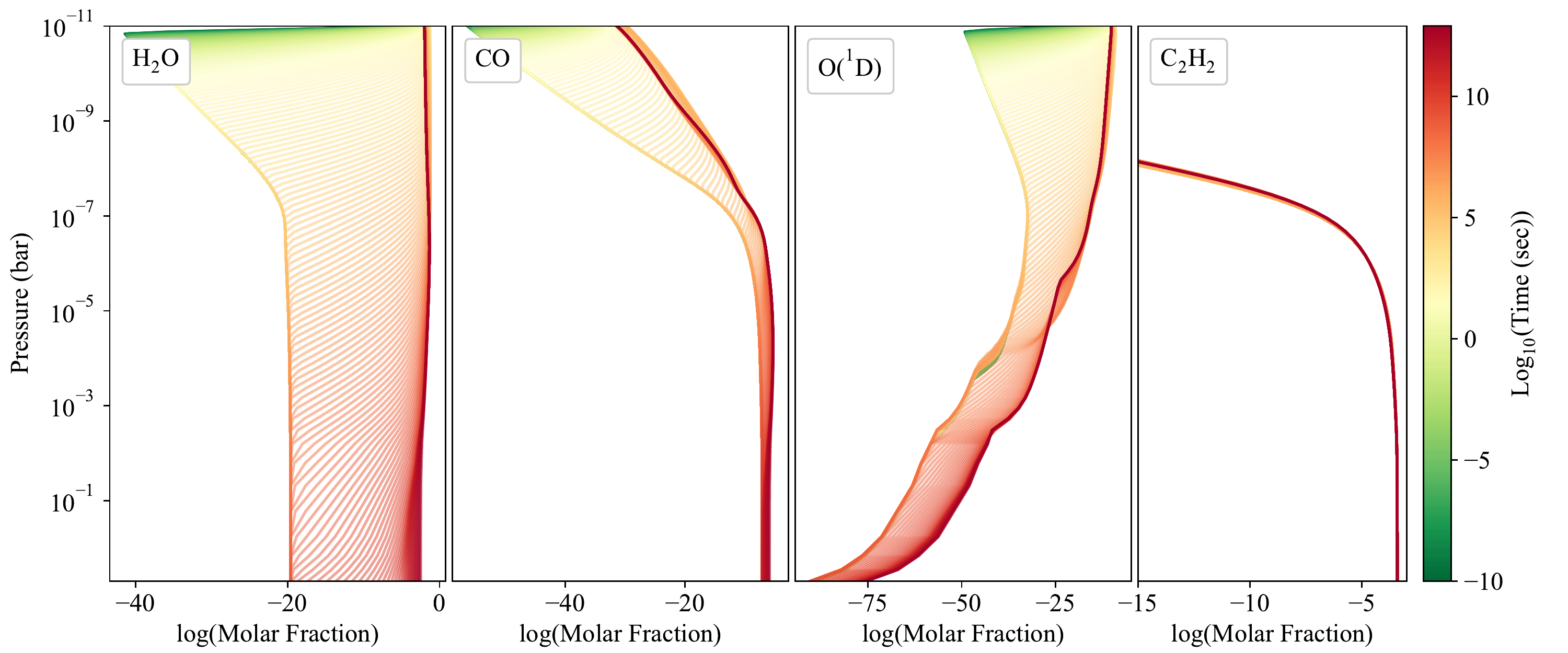}
\caption{ Abundance variations due to $2\times10^5$\;molecules\;cm$^{-2}$s$^{-1}$ water influx at TOA. The model has a temperature structure similar to the Neptune model presented in \hyperref[fig:Neptunecond]{Figure}\;\ref{fig:Neptunecond}, but we initialize the atmospheric composition by its ``photo-diffusion equilibrium''; to study the abundance deviations only due to the water-influx.\label{fig:Neptuneinfluxwater}}
\end{figure*}

First, we only include an influx of \ce{H2O} equal to $2\times10^5$\;molecules\;cm$^{-2}$s$^{-1}$ \citep[as discussed in the ][]{moses_seasonal_2018} to examine how abundances would change. We initialize the atmospheric composition by its ``photo-diffusion equilibrium''\footnote{The steady state after the inclusion of photochemistry and diffusion in the model}. We use the same K\textsubscript{zz} and irradiation spectrum as in the \citet{moses_seasonal_2018} Neptune model. Any deviation from this state will be caused by the imposed influx. \hyperref[fig:Neptuneinfluxwater]{Figure}\;\ref{fig:Neptuneinfluxwater} shows the time evolution of \ce{H2O}, \ce{CO}, O($^1$D) and \ce{C2H2} abundances. As one might expect, \ce{H2O}, \ce{CO}, and O($^1$D) abundances increase, but \ce{C2H2} remains almost the same, due to the lack of free radical carbon. Diffusion plays a pivotal role and transports the deposited material at TOA to the deeper parts of the atmosphere. In this model, the abundance variations caused by \ce{H2O} influx at $10^{-11}$\;bar typically require $10^{5}$-$10^{7}$\;sec (weeks to months) to reach $10^{-7}$\;bar, where the highest altitude measurements of hydrocarbon mixing ratios have taken place \citep[e.g.][]{moses_seasonal_2018}. This time scale would be very different if the required oxygen to explain the observations were supplied by a large cometary impact \citep[e.g.][]{lellouch_dual_2005,luszcz-cook_constraining_2013,moses_seasonal_2018}. These possibilities can be addressed by JWST measurements of Neptune's composition \citep[e.g.][]{roman_spatially-resolved_2018,fletcher_vlt/visir_2018} or a dedicated mission to ice-giants \citep[e.g.][]{hofstdater_ice_2017}.

%%%%%%%%%%%%%%%%%%%%%
%%%%%%%%%%%%%%%%%%%%%

\begin{figure*}
\includegraphics[width=\textwidth]{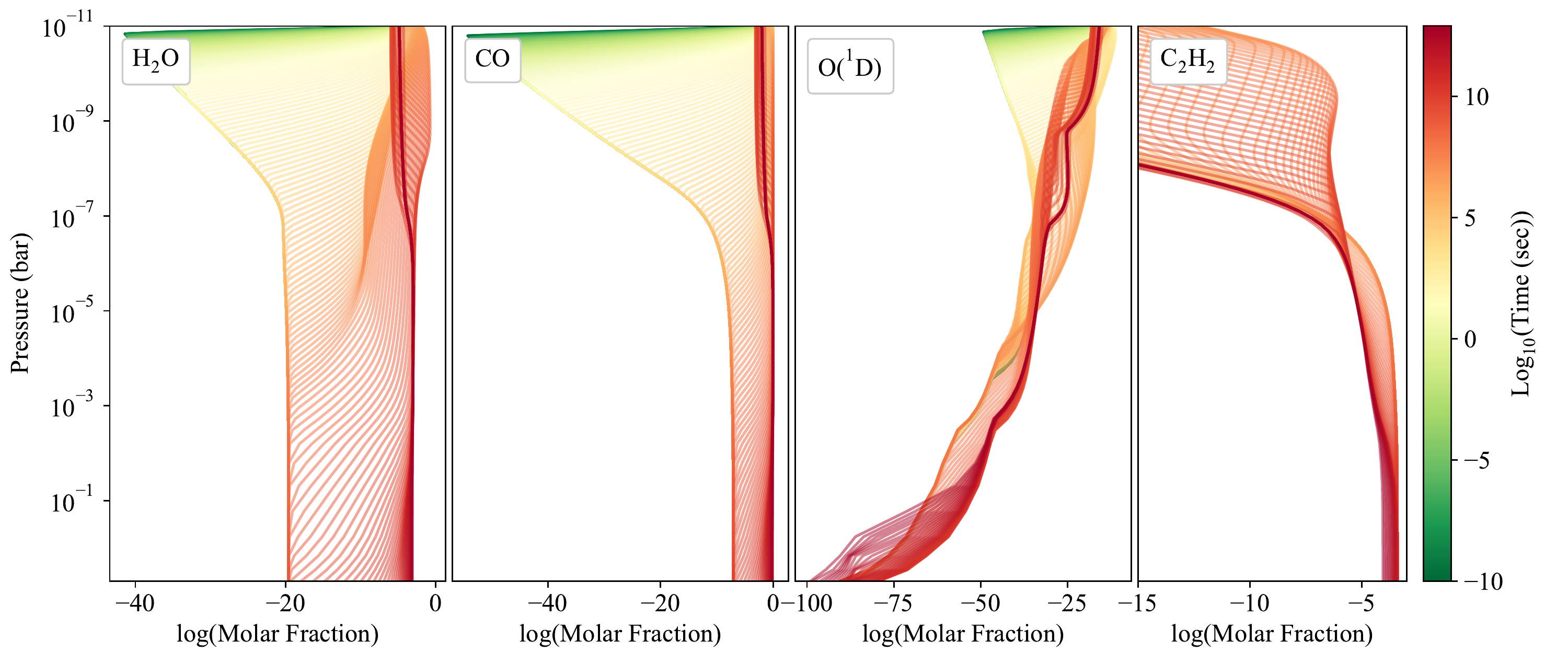}
\caption{Similar to \hyperref[fig:Neptuneinfluxwater]{Figure}\;\ref{fig:Neptuneinfluxwater} but additionally $2\times10^8$\;molecules\;cm$^{-2}$s$^{-1}$ CO influx is included. \label{fig:NeptuneinfluxwaterCO}}
\end{figure*}

In the next case study, we run the same model but include an influx of $2\times10^8$\;molecules\;cm$^{-2}$s$^{-1}$ \ce{CO} \citep[see ][for the motivation]{moses_seasonal_2018} in addition to \ce{H2O}. Under these circumstances, excessive production of free carbon radicals and therefore additional carbon-bearing compounds can be expected. However, individual species at different altitudes might respond to these free radicals differently. Photolysis of incoming \ce{H2O} and \ce{CO} results in the production of free radical H, C, and O; hence an increase in the abundance of their elements is expected (e.g. see the variation of O($^1$D) abundances in \hyperref[fig:NeptuneinfluxwaterCO]{Figure}\;\ref{fig:NeptuneinfluxwaterCO}), unless there are sink terms with higher loss rates. As expected, the production of \ce{CO} and \ce{C2H2} are enhanced early in the simulation, however, at later times (t$\gtrsim10^6$\;sec), \ce{C2H2} abundances seem to be restored to their initial values. This is, of course, not a universal trend and an examination of \ce{C2H2} abundance reveals its depletion at pressures between 0.1 and $10^{-7}$\;bar. Although CO photodissociation enhances \ce{C2H2} production by offering additional free carbons, it also shields it at longer timescales and returns the \ce{C2H2} abundances at TOA to their initial values. This extra shielding also causes the depletion of \ce{C2H2} at mid-pressures and depletion of \ce{H2O} and O($^1$D) abundances at TOA.

%%%%%%%%%%%%%%%%%%%%%
%%%%%%%%%%%%%%%%%%%%%

Similar to these models, boundaries with fixed molar fractions and vertical profile of influxes of reagents (i.e. production rates) can be easily included in \texttt{ChemKM}'s modeling setup. In the case of GCRs, \texttt{ChemKM} includes a separated module to take into account the effect of Galactic/solar cosmic rays. Vertical profiles of production rates of each species with their branching and yields for each GCR reaction can be provided in the current version of the code (similar to the setup of photolysis module).

\bibliographystyle{aasjournal}
\bibliography{references}{}



\end{document}